\documentclass[11pt]{article}
\usepackage{epsf,amsfonts,amsthm,fontenc,indentfirst, delarray,amsfonts,amsmath,amssymb,graphicx}
\DeclareGraphicsExtensions{ps,eps,ps.gz,bmp}

\makeatletter

\def\@citex[#1]#2{\if@filesw\immediate\write\@auxout
        {\string\citation{#2}}\fi
\def\@citea{}\@cite{\@for\@citeb:=#2\do
        {\@citea\def\@citea{,}\@ifundefined
        {b@\@citeb}{{\bf ?}\@warning
        {Citation `\@citeb' on page \thepage \space undefined}}
        {\csname b@\@citeb\endcsname}}}{#1}}
\newif\if@cghi
\def\cite{\@cghitrue\@ifnextchar [{\@tempswatrue
        \@citex}{\@tempswafalse\@citex[]}}
\def\citelow{\@cghifalse\@ifnextchar [{\@tempswatrue
        \@citex}{\@tempswafalse\@citex[]}}
\def\@cite#1#2{{\if@cghi\unskip$\null^{#1}$\else #1\fi\if@tempswa\typeout
        {warning: optional citation argument ignored: `#2'} \fi}}

\def\@biblabel#1{$\null^{#1}$}
\makeatother

\def\Gg{{\cal G}}
\def\ssquare{{\text{\tiny $\blacksquare$}}}
\newcommand{\norm}[1]{\left\lVert#1\right\rVert}
\newcommand{\abs}[1]{\lvert#1\rvert}

\newcommand{\scl}[2]{\langle#1,#2\rangle}

\newcommand{\suum}[2]{\overset{#2}{\underset{#1}{\Sigma}}}

\def\ee{{\cal E}}
\def\cc{{\mathbb{C}}}
\def\rr{{\mathbb{R}}}

\def\ii{{\mathbb{I}}}
\def\aa{{\cal A}}

\def\ll{{\cal L}}

\def\hh{{\cal H}}

\def\oo{{\cal O}}
\def\ss{{\cal S}}

\def\ot{\otimes}

\newtheorem{lem}{Lemme}[section]

\newtheorem{prop}[lem]{Proposition}

\newtheorem{defi}[lem]{Definition}

\def\dm{\lp\begin{array}}
\def\fm{\end{array}\rp}
\def\dbb{\lb\begin{array}}
\def\fbb{\end{array}\rb}
\def\dbn{\left.\begin{array}}
\def\fbn{\end{array}\right.}

\def\ee{{\cal E}}
\def\lb{\left[}
\def\rb{\right]}
\def\lp{\left(}
\def\rp{\right)}

\def\cc{{\mathbb{C}}}
\def\rr{{\mathbb{R}}}

\def\ii{{\mathbb{I}}}
\def\aa{{\cal A}}

\def\ll{{\cal L}}

\def\hh{{\cal H}}

\def\oo{{\cal O}}
\def\ss{{\cal S}}

\def\ttt{\mathcal{T}}
\def\La{\Lambda}
\begin{document}

\title{\bf {\Large An algebraic Birkhoff decomposition for the continuous renormalization group}}
\author{Florian Girelli${}^{d}$, Thomas Krajewski${}^{bc}$, Pierre Martinetti${}^{a}$\\[.2cm]
${}^{a}${\em Instituto Superior T\'ecnico, Lisboa.} pmartin@math.ist.utl.pt\\
${}^{b}$ {\em Centre de physique th\'eorique, CNRS Luminy,Marseille}\\
${}^{c}$ {\em Universit\'e de Provence, Marseille.} thomas.krajewski@cpt.univ-mrs.fr\\
${}^{d}${\em Perimeter Institute, Waterloo.}
fgirelli@perimeterinstitute.ca}
%\date{\small\today}
\maketitle

\begin{abstract}
This paper aims at presenting the first steps towards a
formulation of the Exact Renorma\-lization Group Equation in the
Hopf algebra setting of Connes and Kreimer. It mostly deals with
some algebraic preliminaries allowing to formulate perturbative
renormalization within the theory of differential equations. The
relation between renormalization, formulated as a change of
boundary condition for a differential equation, and an algebraic
Birkhoff decomposition for rooted trees is explicited.
\end{abstract}

\section{Introduction}

During the last five decades, renormalization group theory has
proven to be a major discovery in theoretical physics whose
applications range from high energy physics. its original
birthplace\cite{peterman}, to statistical physics and dynamical
systems, thanks to the work of K. Wilson\cite{kogut}. Originally
proposed as a computational device in quantum field theory
allowing to compare physical theories defined at different energy
scales, it finally turned out to have a deep conceptual
significance. Indeed, it is known since more than twenty years
that renormalization group arguments allow to define the theory in
the sense that one can construct a finite renormalized quantum
field theory by asking it to fulfil a differential equation known
as the Exact Renormalization Group Equation\cite{polchinski}.

On the other side, renormalization recently triggered a couple of
mathematical works \cite{ck} that focused on algebraic aspects of
the substraction procedure and of the resulting renormalization
group invariance. While these works mostly focus on the BPHZ
procedure formulated within the minimal substraction procedure in
dimensional regularization (see for instance ref.[\citelow{gkm}]
for an application to the renormalization  of the wave function),
it is obvious that the framework proposed by Connes and Kreimer is
versatile enough to encompass the ERGE.

This paper aims at presenting the first steps towards a
formulation of the ERGE in the Hopf algebra setting. It mostly
deals with some algebraic preliminaries allowing to formulate
perturbative renormalization within the theory of differential
equations.

In the first part, we present some general results on rooted
treees and their interpretation as a coefficients of formal power
series of non linear operators, in analogy with the theory
developped in numerical analysis under the name of B-series
\cite{butcher}'\cite{wanner}. In the second part, we explicit the
relation between renormalization, formulated as a change of
boundary condition for a differential equation, and an algebraic
Birkhoff decomposition for rooted trees.

While this paper mostly presents some elementary facts, a more
thorough survey involving the precise relation between trees and
Feynman diagrams and their use as computational tools for
effective actions will be the subject of a forthcoming
publication.

\section{Trees and power series of non linear operators}

This section presents the algebraic tools for the interpretation
of the continuous renormalization group in term of Birkhoff
decomposition. We introduce the well known Hopf algebra $H$ of
rooted trees to generate formal power series of (non linear)
operators. Recalling the isomorphism between the group of
characters of $H$ and series with non zero constant term, we focus
on two particular series: the geometrical serie $f_{\phi_1}[X]$
and the exponential serie $f_{\phi_e}[X]$.

\subsection{The Hopf algebra of rooted trees}
Most of the material is detailed in ref.[\citelow{jgb}] and refers
to the original work ref.[\citelow{ck}].  A rooted tree is a
distinguished vertex, the {\it root}, together with a set of
vertices and non intersecting oriented lines. Any vertex has one
and only one incoming line, except the root which has only
outgoing lines. The {\it fertility} of a vertex is the number of
its outgoing lines, its {\it length} is the number of lines of the
(unique) path that joins it to the root.  Two rooted trees are
isomorphic if the number of vertices with given length and
fertility is the same for all possible choices of lengths and
fertilities.  Symbols $T$, one for each isomorphism class,
together with a unit $1$ corresponding to the empty tree generate
a complex commutative algebra $H$ with disjoint union as a
product.

A {\it simple cut} $c$ of a tree $T$ is a (non-empty) subset of
its lines (selected for deletions) such that the path from the
root to any other vertex includes at most one line of $c$.
Deleting the cut lines produces $\text{Card}(c) + 1$ subtrees: the
{\it trunk} $R_c(T)$ which contains the original root and the set
$P_c(T)$ of the pruned branches.  The set of simple cuts of $T$ is
written $C(T)$.
\begin{figure}[h!]
\begin{center}
\begin{tabular}{ccc}
\includegraphics[width=0.3cm]{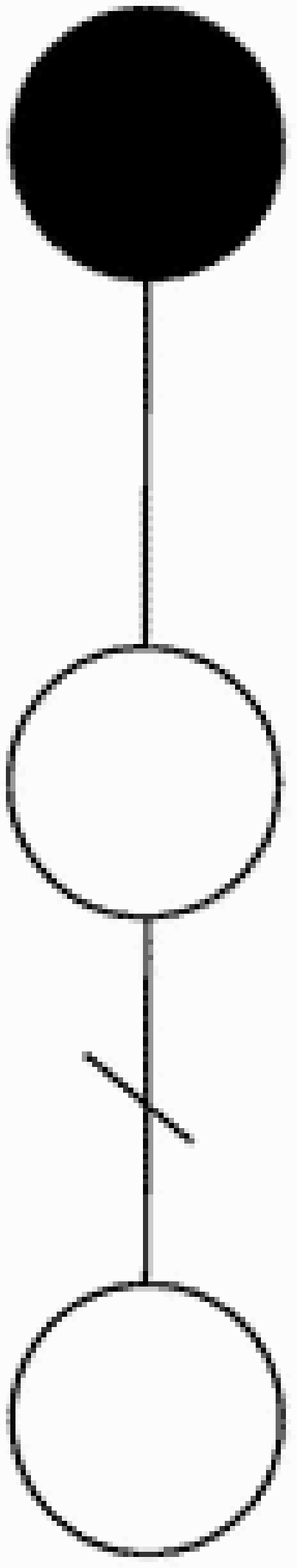} & \includegraphics[width=0.8cm]{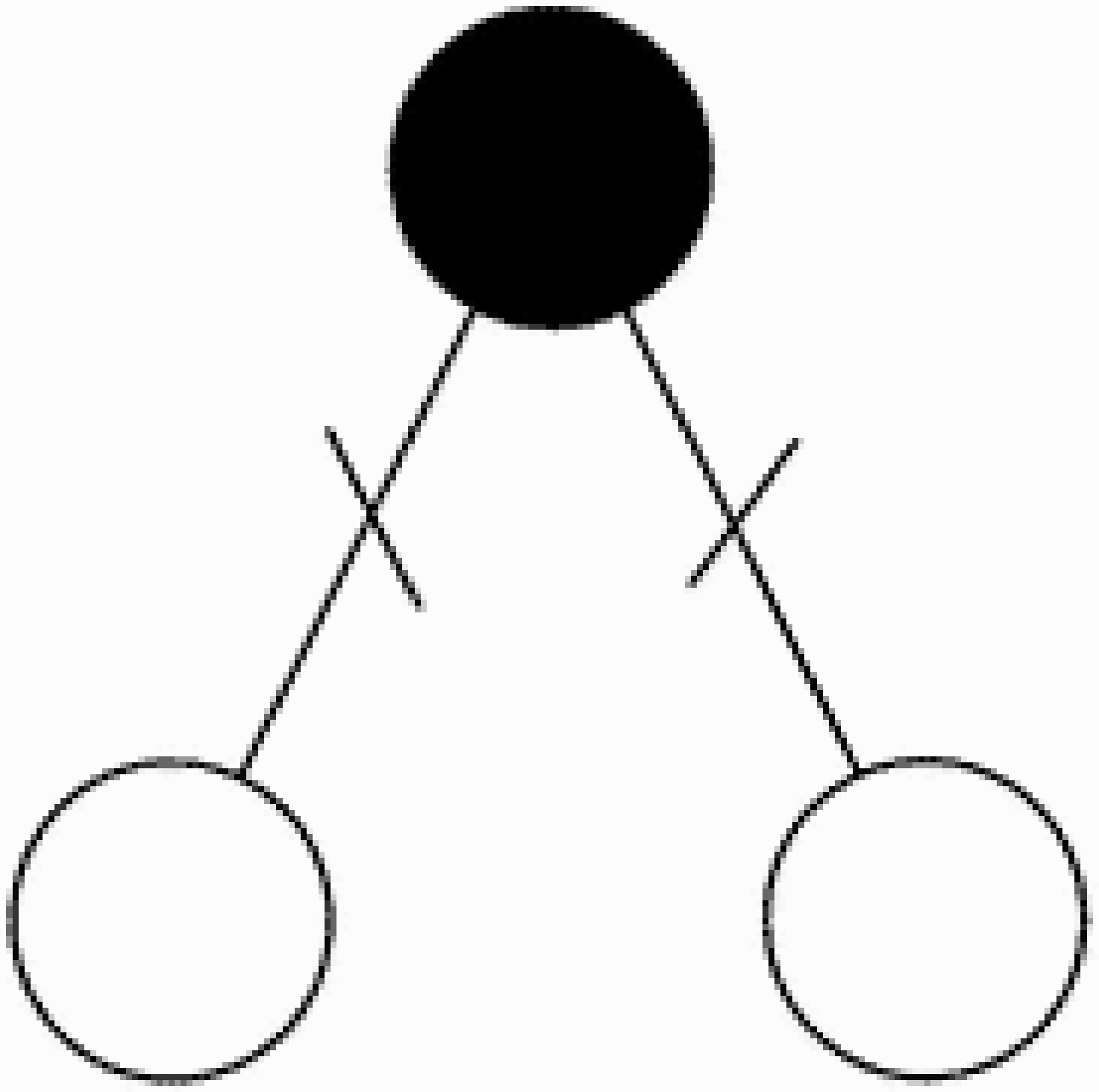}& \includegraphics[width=0.3cm]{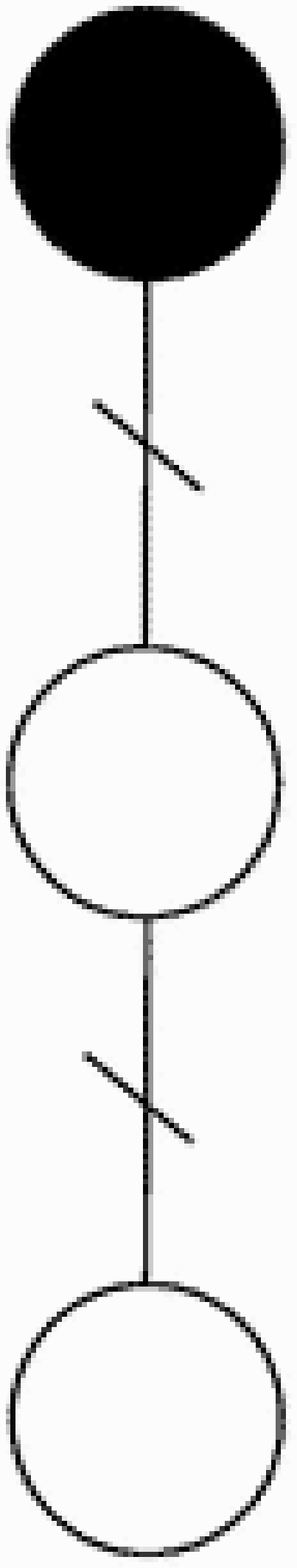}
\\
simple cut & simple cut & non simple cut
\end{tabular}
\end{center}
\end{figure}
{\defi $H$ is an Hopf algebra with counit $\epsilon = 0$ except
$\epsilon(1)=1$, the antipode
\begin{eqnarray}
\label{antipode0}
          S:&     \bullet \mapsto&  - \bullet\\
           \nonumber
            &T \mapsto & - T - \suum{c\in C(T)}{} -  S(P_c(T))
            R_c(T)
\end{eqnarray} and the coproduct
\begin{equation}
\label{coproduit0}
               \Delta(T) = T\otimes 1 + 1\otimes T + \suum{c\in C(T)}{} P_c(T) \otimes
               R_c(T), \quad \Delta(1) = 1\otimes 1.
\end{equation}}
\begin{figure*}[h!]
\centerline{\hspace{0.5cm} \( \Delta
\includegraphics[width=0.8cm]{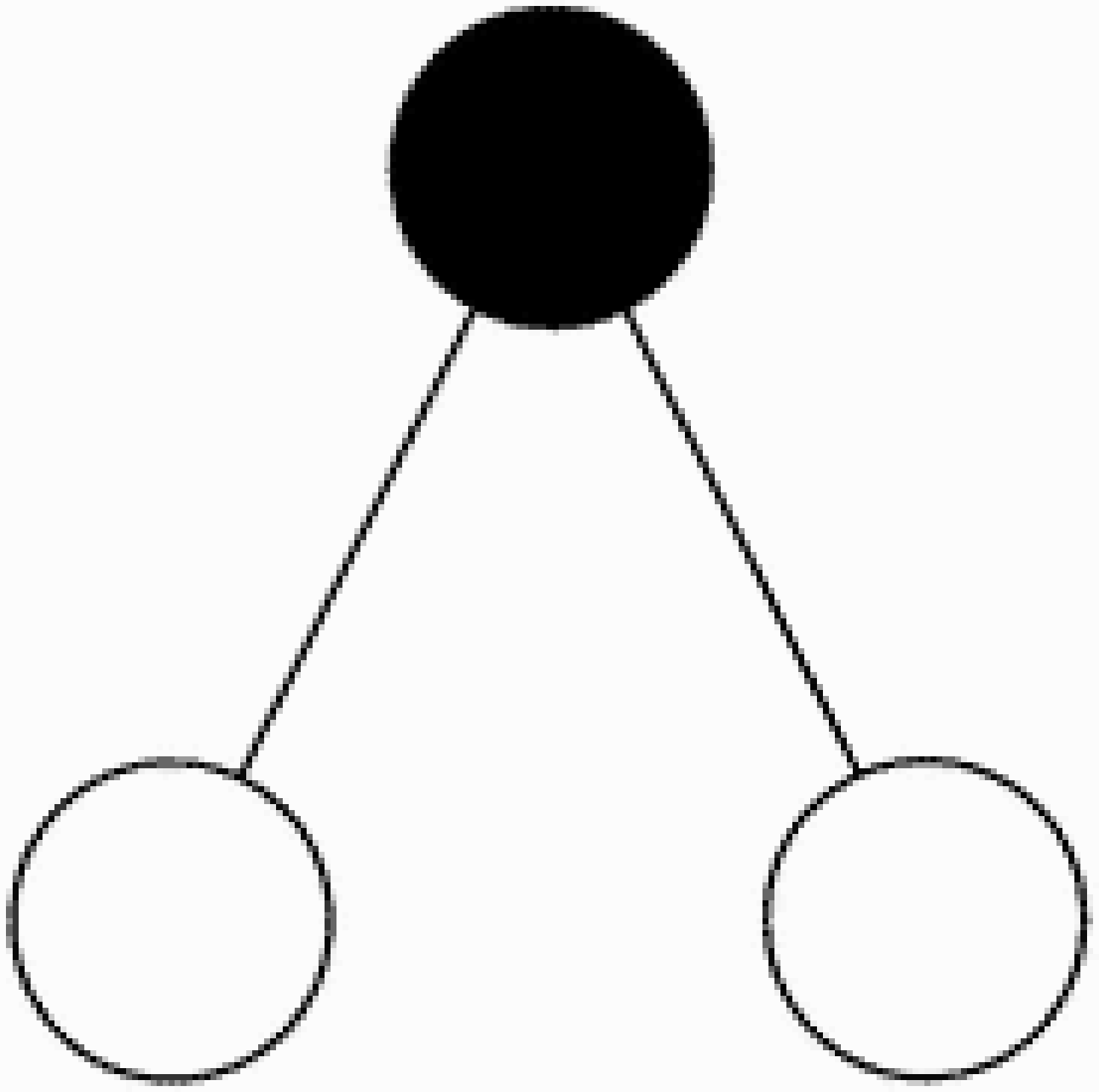}=1\ot
\includegraphics[width=0.8cm]{t32.eps}+
\includegraphics[width=0.8cm]{t32.eps}\ot 1 + 2
\includegraphics[width=0.3cm]{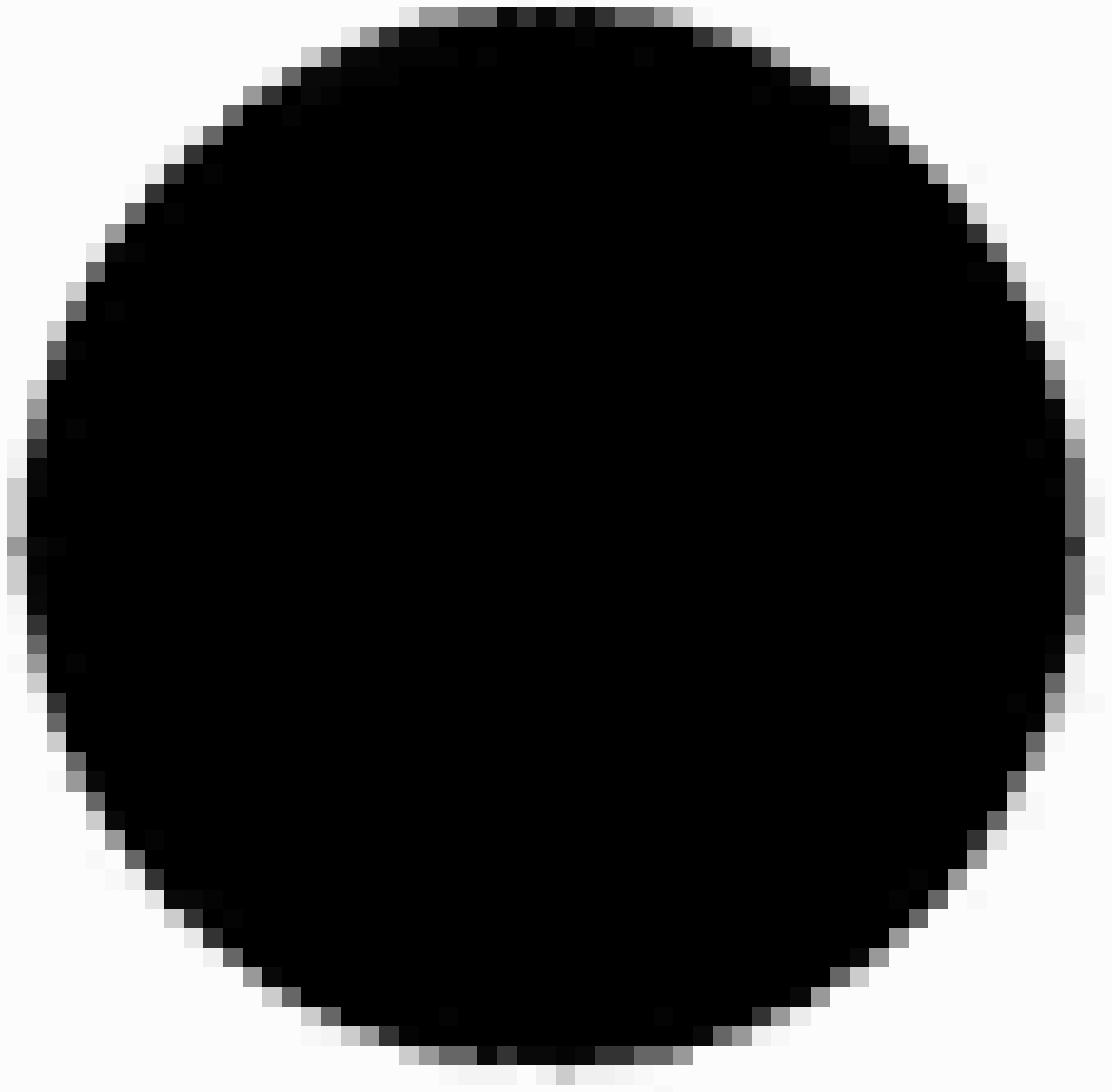} \ot
\includegraphics[width=0.3cm]{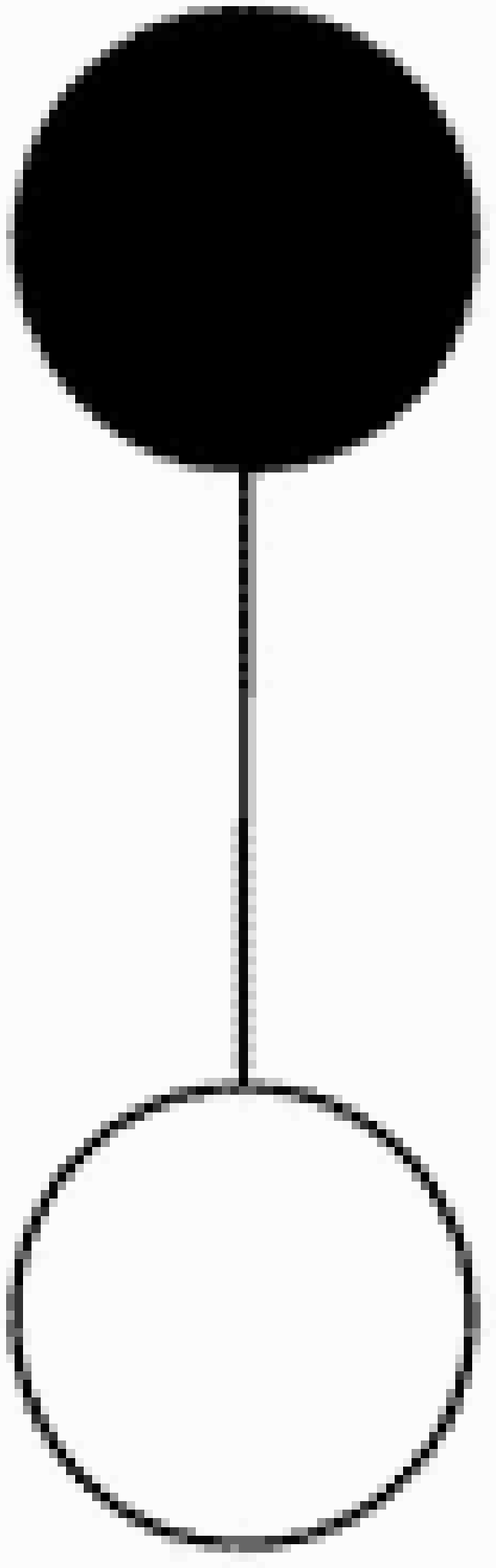}+
\includegraphics[width=0.3cm]{t1.eps}
\includegraphics[width=0.3cm]{t1.eps}\ot
\includegraphics[width=0.3cm]{t1.eps} \)}
\caption{Example of coproduct.}
\end{figure*}

 Among the common quantities associated to a tree $T$, we
use the {\it symmetry factor} $S_T$ which is the number of
isomorphic trees that can be generated by permutation of the
branches, the number of vertices $\abs{T}$ (including the root)
and the {\it depth} which is the maximum length of the vertices of
$T$. The number of vertices is a natural graduation which makes
$H$ a graded connected Hopf algebra \cite{kastler}.
\newline

The dual $H^*$(i.e. the set of linear applications from an Hopf
algebra $H$ to $\cc$) is an algebra with the convolution product
$*$
\begin{equation}
\label{produit}
f*g(a) \doteq \scl{f\otimes g}{\Delta a}
\end{equation}
for $f,g\in H^*$, $a\in H$ and $\scl{}{}$ the duality pairing. For
those $f$ in $H^*$ for which the transposition of the
multiplication $m$ of $H$ takes its values in $H^*\otimes H^*$,
one writes $\Delta (f) \doteq {^tm}(f)$, i.e.
\begin{equation}
\label{coproduit} \Delta f(a\otimes b) = f(ab)\quad\forall a,b\in
H.
\end{equation}
A {\it derivation} is an element $\delta$  of $H^*$ satisfying
\begin{equation}\Delta \delta = \delta\otimes \epsilon + \epsilon\otimes \delta.\end{equation}
 The algebra generated by the derivations and
$\epsilon$ is an Hopf algebra \cite{kastler} $H_*$ for the product
(\ref{produit}) and coproduct (\ref{coproduit}) (antipode, unit
and counit are obtained by transposition of antipode and unit of
$H$). A {\it character} $\phi$ (algebra morphism from $H$ to
$\cc$) is a grouplike element of $H_*$
\begin{equation}\Delta \phi = \phi \otimes \phi.\end{equation}
Characters with product (\ref{produit}) and unity $\epsilon$ form
a group $G$. The inverse is given by the antipode
\begin{equation}
\label{inverse} \phi^{-1} = \phi\circ S.
\end{equation}

When $H$ is graded, with degree $\abs{.}$, derivations are called
{\it infinitesimal characters} because
\begin{equation} \label{expdelta} e^\delta \doteq
\suum{n=0}{+\infty}\; \frac{\delta^n}{n!} \end{equation}
 (well defined since
$\delta^n(a)$ vanishes as soon as $n>\abs{a}$) is grouplike
\cite{jgb}. The correspondence between characters and
infinitesimal characters is one to one:
\begin{equation}
\label{ln} ln\, \phi \doteq \suum{n=1}{+\infty}
\frac{(-1)^{n-1}}{n} (\phi -\epsilon)^n, \; \phi\in G
\end{equation}
is a derivation \footnote{\noindent Identify
$\left\{\begin{array}{ccc} \Delta(ln\,\phi) =
\suum{n=1,p=0}{n=+\infty,p=n}\frac{(-1)^{p+1}}{n}C^n_p\phi^p\ot\phi^p
 &\text{to}&
 ln\, xy= \suum{n=1,p=0}{n=+\infty, p=n} \frac{(-1)^{p+1}}{n}C^n_p
x^py^p,\; x,y \in \rr \\
ln\, \phi \ot \epsilon + \epsilon\ot ln\, \phi =
\suum{n=1}{+\infty}\frac{(-1)^{n-1}}{n} (\phi^n \ot\epsilon +
\epsilon\ot\phi^n) &\text{to}& ln\, x + ln\, y =
\suum{n=1}{+\infty} \frac{(-1)^{n-1}}{n}(x^n +
y^n)\end{array}\right.$}
 satisfying $e^{ln\, \phi} = \phi.$
\footnote{Developing $e^{ln\, \phi} (a)$ in $\tilde{\phi}\doteq
\phi -\epsilon$, terms in $\tilde{\phi}^p$ cancel by combinatoric
for $2\leq p\leq \abs{a}$. For $p>\abs{a}$, $\tilde{\phi}^p(a)$
vanishes because for graded connected Hopf algebra \cite{jgb}
$\Delta(a) = a\ot 1 + 1\ot a + \sum a'\ot a''$ with $\abs{a'},
\abs{a''} < \abs{a}$.}
 Similarly $ln\,
e^\delta$ coincides with $\delta$. The set of infinitesimal
characters is linearly spanned by derivations $Z^T$ that cancel
everywhere but on a given generator
\begin{equation}
\label{z} Z^T(T') \doteq 0 \text{ for any } T'\neq T, \quad
Z^{T}(T) \doteq 1, \quad Z^{T}(1) = 0. \end{equation}

\subsection{Power series of operators}
Let $\ee$ be a Banach vector space with norm $\norm{.}$,
$\ss(\ee)$ the set of smooth applications from $\ee$ to $\ee$ and
$\ll_n^s(\ee)$ the set of $n$-linear symmetric applications from
$\ee^n$ to $\ee$. Take $X\in\ss(\ee)$. For any $x\in\ee$, there
exists an infinite sequence of smooth
\begin{equation}
X^{[n]}_x\in \ll_n^s(\ee)
\end{equation}
such that for any $y$ in the neighborhood of $x$
\begin{equation}X(x+y) = X(x) + X'_x(y) + ...+ \frac{1}{n!}X^{[n]}_x(y,...,y) + \oo(\norm{y}^{n+1}).\end{equation}
For instance when the norm is associated to real coordinates over
a base ${e_{\mu}}$, $X$ is a collection of smooth functions
$X^{\mu}$ with derivatives $X^\mu_{,\nu ...}$ and (summing on
repeated indices)
\begin{equation}
\label{notrg}
X'_x (y) = X^\mu_{,\nu} (x) y^{\nu}e_\mu\;,\;
X''_x(y_1,y_2)\doteq
 X^\mu_{,\nu\rho} (x) \,y_1^\nu\, y_2^\rho\, e_\mu.
\end{equation}
Here we intend to work with a coordinate free notation and we view
$X^{[n]}$ as an element of $\ll^s_n(\ss(\ee))$
\begin{eqnarray}
X^{[n]}(X_1,X_2,..., X_n): &\ee \rightarrow& \ee\\
&x\mapsto& X^{[n]}_x\lp X_1(x), X_2(x),..., X_n(x) \rp.
\end{eqnarray}
One defines a formal power serie in $\hbar$ by Taylor expanding
$X$ around the identity
\begin{equation}
\label{taylor0}
 X(x+\hbar Y(x)) = X(x) + \hbar X'(Y)(x) +
\frac{\hbar^2}{2} X''(Y,Y)(x) + ...
\end{equation}
with $Y\in\ss(\ee)$. When $Y = X$, (\ref{taylor0}) can be written
in a nice graphic way by using rooted trees. Explicitly one
defines
\begin{eqnarray}
X^{\includegraphics[width=0.3cm]{t1.eps}} \doteq X &,& X^{\includegraphics[width=0.3cm]{t2.eps}} \doteq X'(X) \\
X^{\includegraphics[width=0.5cm]{t32.eps}} \doteq \frac{1}{2}
X''(X,X) &,& X^{\includegraphics[width=0.5cm]{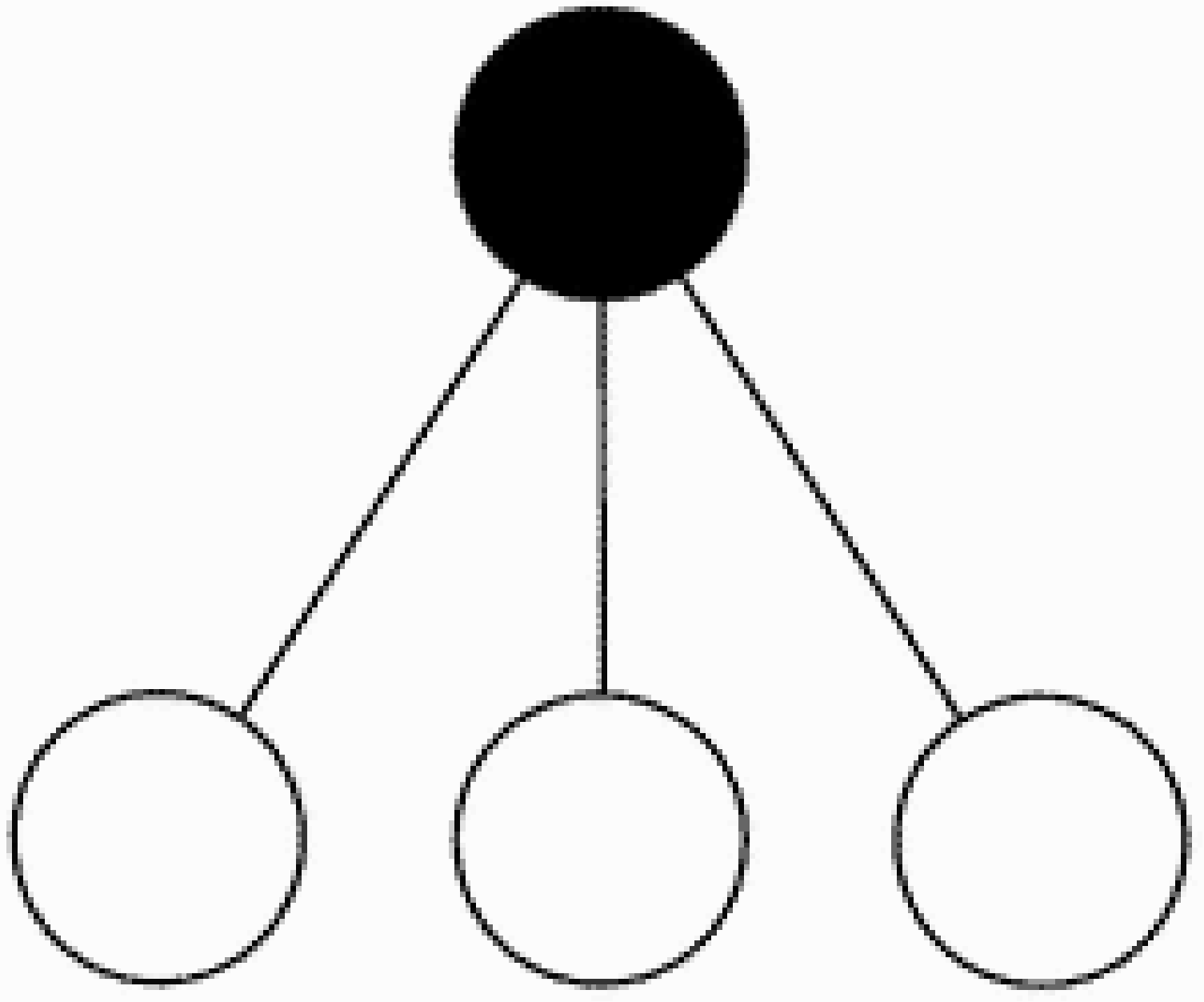}}\doteq
\frac{1}{3!} X'''(X,X,X)
\end{eqnarray}
so that (\ref{taylor0}) writes
\begin{equation}
\label{taylor} X(\ii + \hbar X)=
X^{\includegraphics[width=0.3cm]{t1.eps}}+ \hbar
X^{\includegraphics[width=0.3cm]{t2.eps}}+\hbar^2
X^{\includegraphics[width=0.5cm]{t32.eps}} +\hbar^3
X^{\includegraphics[width=0.5cm]{t43.eps}}    +\hdots
\end{equation}
Formally there is no reason to limit to trees of depth 1 and one
associates to any $T$ the operator recursively defined by
\begin{eqnarray}
\label{xt}
 &X^{T} \doteq \frac{1}{\prod
\theta_i!}X^{[n]}(\underbrace{X^{T_1}, X^{T_1}, ...}_{\theta_1
\text{ times}}, \underbrace{X^{T_2}, ...}_{\theta_2\text{ times}},
..., \underbrace{X^{T_m}, ...}_{\theta_m\text{ times}}),& \\
&X^{1}=\ii&
\end{eqnarray}
where the $T_i$'s are the subtrees of $T$ obtained by promoting as
roots the vertices of $T$ of length $1$ and $n = \sum_{i=1}^{m}
\theta_i$ is the fertility of the root of $T$. For instance
\begin{equation}
X^{\includegraphics[width=0.3cm]{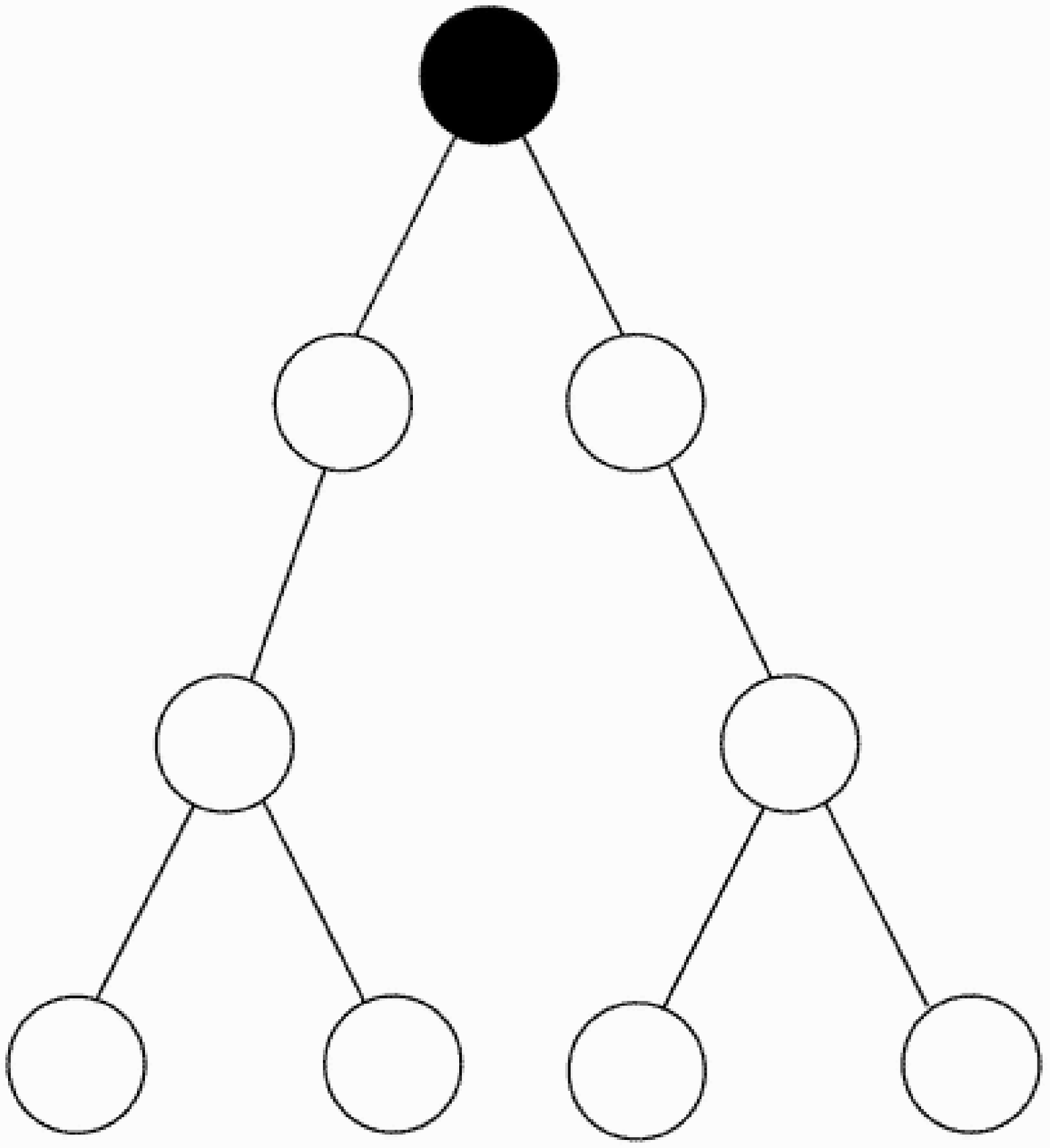}} =
\frac{1}{8}X''(X'(X''(X,X)), X'(X''(X,X))).\end{equation} Note
that the numerical factor in $X^T$ is $\frac{1}{S_T}$. In the
following, we use the shorthand notation
\begin{equation}
\label{shorthand}
 X^{T} =\frac{1}{\prod
\theta_i!}X^{[n]}(X_i^{\theta_i}).
\end{equation}
When $X$ is a linear operator, $X^T = 0$ as soon as $n
> 1$. The following construction, inspired from the theory of
Butcher group \cite{wanner}, is mostly interesting for non linear
operators and provide a generalization of Taylor expansion
(\ref{taylor}) by associating to any smooth $X$ a formal power
serie $f[X]$
\begin{equation}
\label{fx} f[X] \doteq \sum_{T} f_T X{^T} \hbar^{\abs{T}}
\end{equation}
where $f_T$ are complex numbers and the sum runs over all
generators (including unit $1$).

The set $\Gg$ of series with $f_1 =1$ is in one to one
correspondence with the characters of $H$. Moreover $\Gg$ with the
composition of series
\begin{equation}
\label{composition} f\circ g[X] = \sum_{T} f_T X^T \lp \sum_{T'}
\,g_{T'} X{^{T'}}\hbar^{\abs{T'}}\rp \hbar^{\abs{T}}
\end{equation}
is a group, isomorphic to $G^{op}$, the opposite of the group of
characters of $H$. That $\Gg$ is a group is known from the theory
of Butcher group. However by using the Hopf algebraic structure of
trees we present a proof of the isomorphism
\begin{equation}
\label{eqiso} G^{op} \sim \Gg\end{equation} that does not require
to compute explicitly the coefficients of the product serie
(\ref{composition}), as this is done in ref.[\citelow{wanner}] for
instance. The main tool for our proof is a slight adaptation to
our component free notation of Caley's relation\cite{cayley}
between differentials and trees. It appears indeed that the
differentials of the $X^T$'s are easily computed in a graphic way.
In the simplest cases this is immediate from definition
(\ref{xt}), for instance
\begin{equation}{X^{\bullet}}'(X^{\includegraphics[width=0.3cm]{t32.eps}}) =
X^{\includegraphics[width=0.3cm]{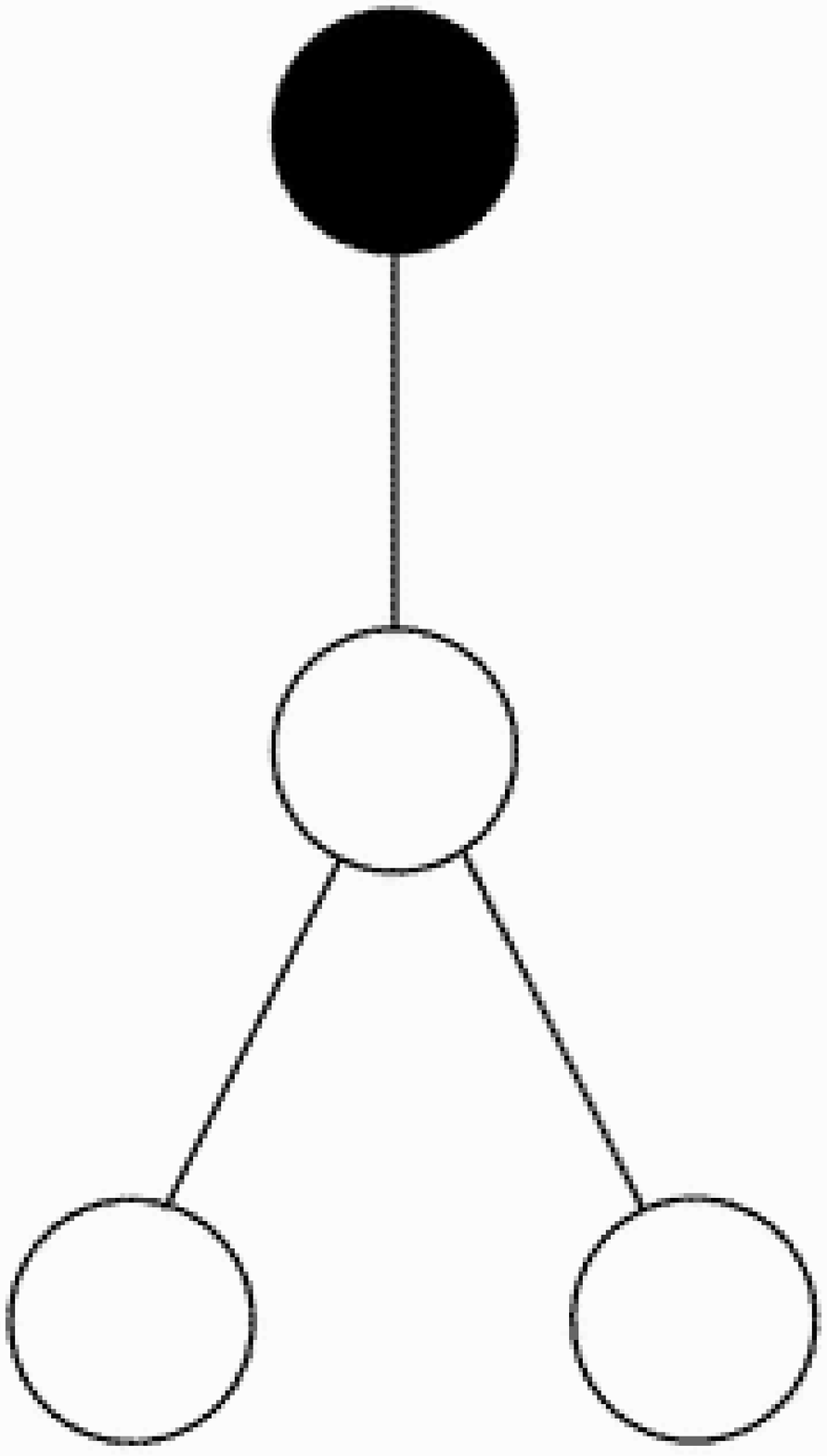}} \,\text{ and }\,
{X^{\bullet}}''(X^{\includegraphics[width=0.3cm]{t32.eps}},
X^{\includegraphics[width=0.3cm]{t32.eps}})=
2X^{\includegraphics[width=0.3cm]{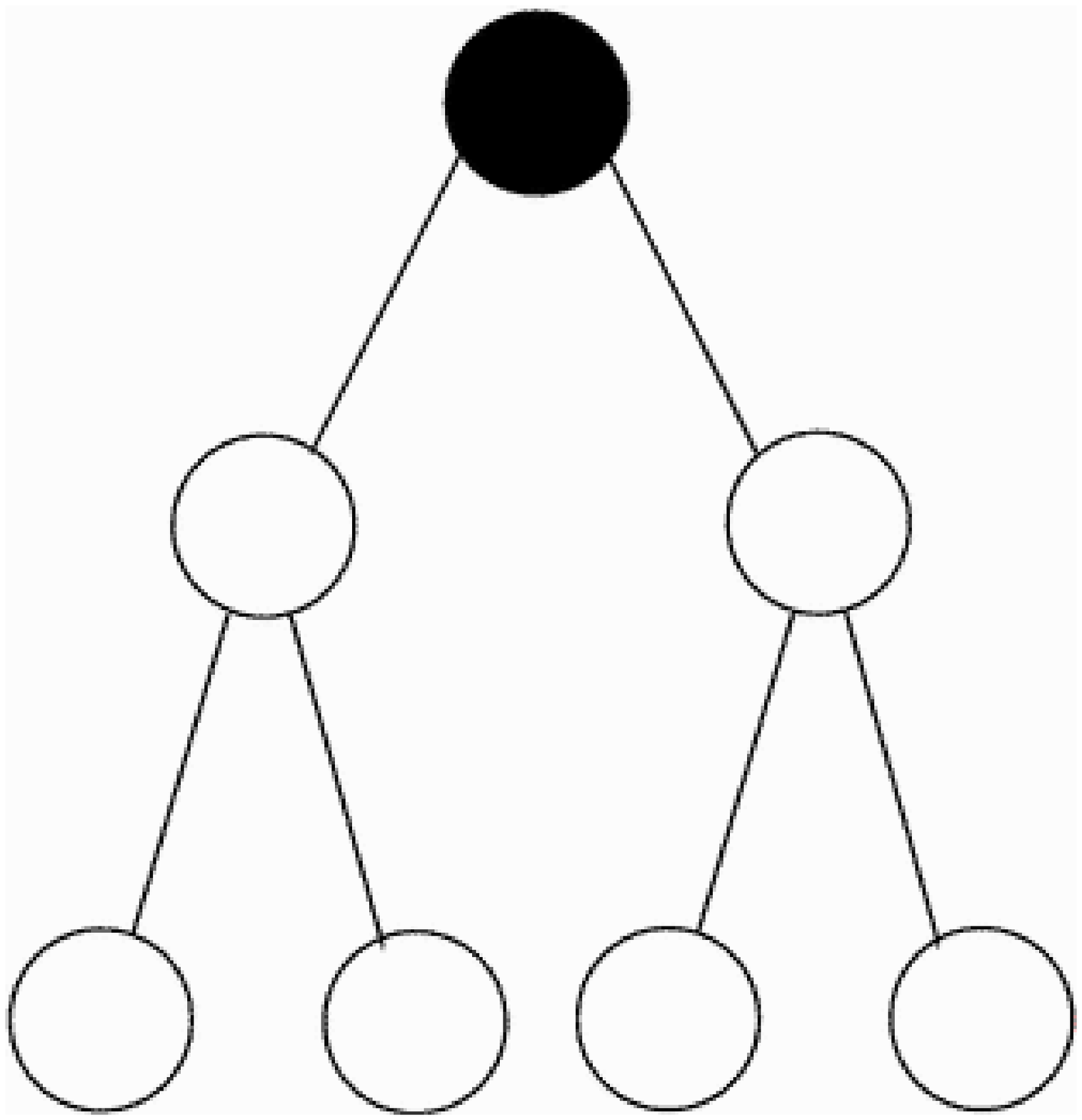}}.\end{equation} The
differentiation process consists in grafting on a tree as many
trees as the order of derivation, with a suitable numerical
ponderation reflecting the symmetry of the argument.

{\lem \label{lemmeun} For any trees $T$, $T_1$, ..., $T_l$ and
integers $\beta_1,... \beta_l$
\begin{equation}X^{T[k]}(\underbrace{X^{T_1}, X^{T_1}, ...}_{\beta_1 \text{ times}}, ...,\underbrace{X^{T_l}, X^{T_l}, ...}_{\beta_l \text{ times}})  =
\Pi\, \beta_i !\sum_{\tilde{T}} n(\tilde{T}; T , \Pi\,
T_i^{\beta_i}) X^{\tilde{T}}
\end{equation}
 where $k\doteq \sum_{i=1}^{l}\beta_i$ and $n(\tilde{T}; T , \Pi\, T_i^{\beta_i})$ is the number of simple cut
 $c$ of $\tilde{T}$ such that \begin{equation}R_c(\tilde{T}) = T\,\text{ and }\,P_c(\tilde{T})= \Pi\, T_i^{\beta_i}.\end{equation}}

\noindent{\it Proof.\;} Let $X\in\ss(\ee)$, $x\in \ee$ and
$y\doteq y_1 + ... + y_J$ in a neighborhood of $x$. For any
collection $\{z_i\}_{i\in[1,n]}$ of elements of $\ee$,
 \footnote{The formula can be obtained by comparing
 \begin{equation}
\label{diff0}X(x + y + z) = X(x+ y) + \sum_{n= 1}^{+\infty}
\sum_{[\alpha_i] = n} \frac{1}{\prod \alpha_i!} X^{[n]}_{x+ y}
(z_i^{\alpha_i})\end{equation}
 where $z = z_1 + ... + z_n$ is in a neighborhood of $x + y$, to
\begin{equation}
\label{diff} X(x + y + z) = X(x) + \sum_{n=1}^{+\infty}
\sum_{[\beta_j, \alpha_i] = n} \frac{1}{\prod \beta_j! \prod
\alpha_i!}X^{[n]}_x (y_j^{\beta_j}, z_i^{\alpha_i}).\end{equation}
 Choosing all $\alpha_i =1$, the term $X^{[n]}_{x+ y} (z_i)$ in (\ref{diff0}) is identified to the sum of terms
 of (\ref{diff}) that are linear in each
 $z_i$. For instance
\begin{equation} \nonumber X'_{x+ y_1 + y_2} (z_1) =
X'_x(z_1) + X''_x(z_1, y_1) + X''_x(z_1, y_2) +
 X'''_x(z_1, y_1, y_2) + \frac{1}{2} X'''_x(z_1, y_1,
y_1) + \frac{1}{2} X'''_x(z_1, y_2, y_2) + ... \end{equation}}
\begin{equation}
\label{xderive} X^{[n]}_{x+ y} (z_i) = \sum_{\beta = 0 }^{+\infty}
\sum_{[\beta_j] = \beta}^{}
  \frac{1}{\prod \beta_j!} X^{[n + \beta]}_x
 (y_j^{\beta_j},z_i)
 \end{equation}
where the second sum runs on all the configurations of the
$\beta_j$'s such that $\sum_{i=1}^{J}\beta_j = \beta$ and we use
notation similar as (\ref{shorthand}). Then
\begin{eqnarray}
 X^T (x+ y) &=& \frac{1}{\prod \theta_i !} X^{[n]}_{x + y} (X_i^{\theta_i}(x+y)) \\
\label{xtt}
            &=&  \frac{1}{\prod \theta_i !}\sum_{\beta = 0
}^{+\infty} \sum_{[\beta_j] = \beta}^{} \frac{1}{\prod \beta_j!}
X^{[n + \beta]}_x (y_j^{\beta_j},X_i^{\theta_i}(x + y)).
\end{eqnarray}
Inserting
\begin{equation}
\label{xti}X_i(x+y) = X_i(x) +
\sum_{r=1}^{+\infty}\sum_{[\beta_j'] = r} \frac{1}{\prod
\beta'_j!} X_{ix}^{[r]}(y_j^{\beta'_j})\end{equation} and choosing
$\{y_j\}$ of cardinality $J=k$, one identifies $X^{T[k]}_x(y_j)$
as the terms of (\ref{xtt}) linear in each $y_j$, namely those
appearing with either $\beta_j$ or $\beta'_j=1$. Explicitly, for a
tree $T$ with depth $1$, \footnote{for instance with notation
(\ref{notrg})
\begin{eqnarray}
{X^{\includegraphics[width=0.3cm]{t32.eps}}}'(X^\bullet) &=&
\frac{1}{2} (X^\mu_{,\nu\rho}X^\nu X^\rho)_{,\delta}X^\delta e_\mu
= \frac{1}{2} X^\mu_{,\nu\rho\delta}X^\nu X^\rho X^\delta e_\mu +
\frac{1}{2} X^\mu_{,\nu\rho}X^\nu_{,\delta} X^\rho X^\delta e_\mu+
\frac{1}{2} X^\mu_{,\nu\rho}X^\nu X^\rho_{,\delta} X^\delta
e_\mu\\
&=& 3X^{\includegraphics[width=0.3cm]{t32.eps}} + X^{
\!\!\!\!\tiny{\begin{array}{c}
\bullet\vspace{-1.7mm}\\
\wedge\vspace{-1.5mm}\\
\circ\,\circ\vspace{-1.7mm}\\ \hspace{1.8mm}\shortmid\vspace{-1.5mm}\\
\hspace{1.8mm}\circ
\end{array}}}
\end{eqnarray}}
\begin{eqnarray*}
X^{T'}(X^{\tilde{T}_1}) &=& \frac{1}{n!}X_x^{[n+1]}
(X^{\tilde{T}_1},X^n) + \frac{1}{(n-1)!}
X_x^{[n]}(X'(X^{\tilde{T}_1}), X^{n-1})\\
&=& \sum_{\tilde{T}} n(\tilde{T}; T , \tilde{T}_1)) X^{\tilde{T}}
\end{eqnarray*}
for any $\tilde{T}_1$. Similarly for any $k$ and distinct $T_j$'s
one obtains
\begin{equation}X^{T[k]}(X^{\tilde{T}_1}, ..., X^{\tilde{T}_k}) = \sum_{\tilde{T}} n(\tilde{T}; T, \Pi\,\tilde{T_j})X^{\tilde{T}}.\end{equation}
When some of the $\tilde{T_j}$'s are identic, the extra factor
$\frac{1}{\beta_j !}$ is absorbed by the definition of $\tilde{T}$
but such terms have to be identified as part of
 \begin{equation}\frac{1}{\beta_j !}X^{T[\beta_j]}(\tilde{X}_j^{\beta_j}).\end{equation}
 Hence the lemma
for any $T$ of depth $1$. The final result is obtained recursively
on the depth of $T$. \hfill $\blacksquare$
\newline

{\prop
 \label{homomorphism}
$\Gg$ is a group, isomorphic to $G^{op}$.}
\newline

\noindent {\it Proof.} Since $\Gg$ is in bijection with $G$, we
just have to show that
\begin{equation} \label{resultat1}
 \sum_{T} \phi(T) X^T \lp \sum_{T'} \psi(T') X^{T'} \hbar^{\abs{T'}}\rp \hbar^{\abs{T}}
=  \sum_{T} (\psi*\phi)(T) X^T \hbar^{\abs{T}}
\end{equation}
for any $\phi, \psi \in G$. In the l.h.s. $\phi$ is evaluated on
generators only. By (\ref{produit}) and (\ref{coproduit0}) one
checks that the same is true for the r.h.s. Moreover both sides
are linear in $\phi$ so this is enough to work with the
infinitesimal character
\begin{equation}
\label{phi}  \phi = Z^{T_0}
\end{equation}
for a fixed $T_0$. Note that a similar simplification is not
available for $\psi$ since $P_c(T)$ may not be a single generator.
Then r.h.s. of  (\ref{resultat1}) reduces to
\begin{equation}
\label{rhs2}  X^{T_0} \hbar^{\abs{T_0}} + \suum{T}{}\; \suum{c\in
\tilde{C}(T)}{}\psi(P_c(T)) X^{T}\hbar^{\abs{T}}
\end{equation}
where $\tilde{C}(T)$ is the set of simple cuts of $T$ such that
$R_{c}(T) = T_0$. The l.h.s of (\ref{resultat1}) developed thanks
to (\ref{taylor}) writes
\begin{equation}
\label{lhs1} X^{T_0} \hbar^{\abs{T_0}} + \sum_{T_1} \psi(T_1)
{X^{T_0}}'(X^{T_1}) \hbar^{\abs{T_0}+\abs{T_1}}  + ... + \sum_{\{
T_i^{\beta_i}\}} \frac{\Pi\, \psi(T_i)^{\beta_i}}{\Pi\,
\beta_i!}{X^{T_0}}^{[n]}(T_i^{\beta_i}) \hbar^{\sum_i \abs{T_i}} +
...
\end{equation}
where sums run over generators distinct from $1$ and we use a
similar notation as in (\ref{shorthand}). Thanks to lemma
\ref{lemmeun}, (\ref{lhs1}) = (\ref{rhs2}). \hfill $\blacksquare$
\newline

 Thanks to this isomorphism one can easily compute some interesting
 series.
 Let us write
\begin{equation} f_{\phi}[X]\end{equation}
the formal power serie associated to a character $\phi$ by the
isomorphism \ref{homomorphism}. We define the {\it exponential
serie} as the one corresponding to the character
\begin{equation}
\phi_e\doteq e^{Z\bullet}
\end{equation}
defined by (\ref{expdelta}) and (\ref{z}). Since (see
ref.[\citelow{ck}] for the definition of the factorial $T!$ of a
tree)
\begin{equation}\phi_e(T) =\frac{1}{T!}\end{equation}
the exponential serie writes
\begin{equation}
\label{expo} f_{\phi_e}[X] = \sum_{T}
\frac{1}{T!}\hbar^{\abs{T}}X^T .\end{equation} Another nice
example is the {\it geometrical serie}
\begin{equation}
\label{seriegeo} (\ii - \hbar X)^{-1}  = f_{\phi_1}[X] = \sum_T
\hbar^{\abs{T}} X^T
\end{equation}
obtained by noting that the constant character
\begin{equation}
\label{phiun}
 \phi_1 : T \mapsto 1 \quad \forall T \end{equation}
 is the inverse of
the character $T \mapsto 0$ except $1\mapsto 1$, $\bullet\mapsto
-1$.
%%%%%%%%%%%%%%%%%%%%%%%%%%%%%%%%%%%%%%%%%%%%%%%%%%%%%%%%
%I
%have also on my transparencies a result on the convergence, we
%could include it -- je ne comprends pas trop ce point dans tes transparents.
%On en a discute avec Florian mais nos interpretations divergent:
%- est ce que tu definis une notion de divergence pour les series ?
%- ou bien est ce que tu verifies que la serie geometrique converge
%pour la definition usuelle de la convergence d'une serie ?( a savoir
% lim n-> \infty \norm{U_n - U}= 0. Avec ici le seul (leger probleme) que la
%notion de T-> infty n'est pas defini mais j'imagine qu'on peut
%prendre \abs{T}->infty).
%%%%%%%%%%%%%%%%%%%%%%%%%%%%%%%%%%%%%%%%%%%%%%%%%%%%%%%%
\section{Fixed point equation and algebraic Birkhoff decomposition}

The geometrical and exponential series (\ref{expo}),
(\ref{seriegeo}) are interesting to solve fixed point equation. In
particular if we view a fixed point equation as the integral form
of a differential equation then the change of initial condition
(corresponding to the change of the constant in the fixed point
equation) is coded into the group $G_2$ of characters of the Hopf
algebra of decorated rooted trees. Namely if $\phi\in G_2$
represents the solution for initial condition $x_0$ and $\phi_+\in
G_2$ corresponds to the solution for initial condition $x_1$, then
the decomposition
\begin{equation}
\label{birkmoinsun}
 \phi_+ = \phi_-
* \phi
\end{equation}
turns out to define an {\it algebraic Birkhoff decomposition} (see
definition \ref{defibirkhoff}) similar as the one introduced by
Connes and Kreimer in their seminal paper ref.[\citelow{ck}].

Using trees to solve fixed point and differential equations is at
the heart of the theory of $B$-series (see ref.[\citelow{brouder}]
for a nice overview of such applications). However, to the
knowledge of the authors, the interpretation in terms of Birkhoff
decomposition is not found in the litterature.

\subsection{Change of initial conditions in fixed point equations}

Let us start by the fixed point equation
\begin{equation}\label{ptfixe}
x= x_0 + \hbar X(x)
\end{equation}
where $X \in \ss(\ee)$, $\hbar$ is small and $x_0\in \ee$.
(\ref{ptfixe}) is formally solved by writing
\begin{equation}\label{ptfixe2}
x = (\ii - \hbar X)^{-1}(x_0)
\end{equation}
 that is to say, thanks to (\ref{seriegeo}),
\begin{equation}\label{ptfixe3}
x= f_{\phi_1}[X](x_0) = \sum_T \hbar^{\abs{T}} X^T(x_0).
\end{equation}
This is the same serie as the one found by recursively developing
$x = x_0 + \hbar X( x_0 + \hbar X(X_0 +...)$ ).

Consider now a curve $x: \rr \rightarrow \ee$ given by
\begin{equation}
\label{equadiff}
 x(t) = x_0+ \hbar \int_{t_0}^t X(x(u))du
\end{equation}
for some $X\in\ss(\ee)$. This is the integral form of the
differential equation
\begin{equation}
\label{equadiff2} \frac{dx}{dt}= \hbar X(x)  \; , \; x(t_0)= x_0.
\end{equation}
By recursively developing
\begin{equation}x(t)= x_0 + \hbar \int_{t_0}^t X\lp x_0+ \hbar \int_{t_0}^{t_1} X(x_0 + \int_{t_0}^{t_2}...)dt_2\rp
dt_{1}
\end{equation}
one finds \cite{brouder}
\begin{equation}
 x(t) = \sum_{T} \frac{1}{T!}(t-t_0)^{\abs{T}}
\hbar^{\abs{T}}X^T (x_0)\end{equation} that is to say, according
to (\ref{expo}),
\begin{equation}
\label{soluexpo} x(t) = f_{\phi_e}[X_0](x_0)\end{equation} where
\begin{equation}
\label{xzero} X_0 \doteq (t-t_0)X.
\end{equation}
One may be tempted to solve (\ref{equadiff2}) formally by writing
\begin{equation}
\label{essai}
 x(t) = e^{\hbar X_0}(x_0).
\end{equation}
The genuine definition of the exponential of an operator
{\footnote{$\text{exp } (X) = \ii + X + \frac{1}{2} X\circ X +
...$}} is not compatible with (\ref{soluexpo}). However {\it
defining}
\begin{equation}e^{\hbar X_0}
\doteq f_{\phi_e}[X_0]
\end{equation}
ensures that (\ref{soluexpo}) equals (\ref{essai}). In other terms
rooted trees provide a definition of the exponential of a (non
linear) operator which is compatible with the resolution of fixed
point equations.

The same differential equation can be directly solved in its
integral form (\ref{equadiff}) by considering the Banach vector
space $\ee'$ of curves in $\ee$ and the operator
$\chi\in\ss(\ee')$ defined by
\begin{equation}\chi(x): t\mapsto \int_{t_0} ^t X(x(u))du.
\end{equation}
 Considering $x_0$ as the curve $t \mapsto x_0$, (\ref{equadiff}) reads as the fixed
point equation
\begin{equation} x =  x_0 + \hbar\chi(x)
\end{equation}
whose formal solution is given by (\ref{ptfixe3})
\begin{equation}\label{chi} x = (\ii - \hbar\chi)^{-1}(x_0) = f_{\phi_1}[\chi](x_0).\end{equation}

 Trees are especially
useful to deal with change of initial condition. Let us fix
\begin{equation}
x(t_1) = x_1
\end{equation}
 for a given $x_1\in\ee$ and $t_1\neq t$. Solution (\ref{soluexpo}) becomes
\begin{equation}
\label{solun}
 x(t) = f_{\phi_e}[X_1](x_1)
\end{equation}
where
\begin{equation}
\label{condinitdeux}
 X_1 \doteq (t-t_1)X.
\end{equation}
 In order to keep a trace of the solution with initial
condition at $t_0$, let us write (\ref{solun}) for $t=t_0$,
\begin{equation}
\label{solzeroun}
 x_0 = f_{\phi_e}[X_1 - X_0](x_1),
\end{equation}
which, inserted in (\ref{soluexpo})
\begin{equation} x(t) = f_{\phi_e}[X_0]\circ f_{\phi_e}[X_1 - X_0])(x_1),
\end{equation}
and compared to (\ref{solun}) yields
\begin{equation}
\label{birkseriemoinsdeux} f_{\phi_e}[X_1] =
f_{\phi_e}[X_{0}]\circ f_{\phi_e}[X_1 - X_0].
\end{equation}
A similar decomposition can be found for the integral form
(\ref{equadiff}) of the differential equation. The solution for
initial condition (\ref{condinitdeux}) is
\begin{equation}
\label{xcondinit}
 x = f_{\phi_1}[\zeta](x_1)
 \end{equation}
with
\begin{equation}
\zeta (x): t\mapsto \int_{t_1} ^{t} X(x(u))du.
\end{equation}
 Factorizing
\begin{equation}
(\ii - \hbar \zeta)^{-1} =  (\ii - \hbar \chi)^{-1} \circ (\ii -
\hbar \xi)^{-1}
\end{equation}
where we define
\begin{equation}
\hbar\xi \doteq \ii - (\ii - \hbar \zeta)\circ(\ii - \hbar
\chi)^{-1},
\end{equation}
yields by (\ref{seriegeo})
\begin{equation}
\label{birkserie} f_{\phi_1}[\zeta] = f_{\phi_1} [\chi] \circ
f_{\phi_1}[\xi].
\end{equation}

Both (\ref{birkseriemoinsdeux}) and (\ref{birkserie}) involves a
unique character $\phi_1$ or $\phi_e$ and three distinct operators
$\chi$, $\zeta$, $\xi$ or $X_0, X_1, X_1 - X_0$. To fix notations
we now focus on the decomposition of geometrical series
(\ref{birkserie}) but the following is also true for exponential
series. In order to take advantage of the isomorphism
(\ref{eqiso}) one needs series involving the same operator. This
can be obtained by considering the group $G_2$ of characters of
the Hopf algebra $H_2$ of {\it decorated rooted trees}, i.e. trees
whose vertices carry a decoration chosen in a set of cardinality
two, in our case of interest either a bullet or a square. To any
generator $\mathcal{T}$ of $H_2$, one associates
$Y^\mathcal{T}\in\ss(\ee')$ defined in a similar manner as in
(\ref{xt}) with the extra rule that bullets are associated to
$\xi$ and squares to $\zeta$. For instance
\begin{eqnarray}
\label{y} &Y^\ssquare
 = \zeta,\quad Y^{\bullet}= \xi&\\
&Y^{ \!\!\!\!\tiny{\begin{array}{c}
\ssquare\vspace{-1.6mm}\\
\wedge\vspace{-1.5mm}\\
\circ\,\circ
\end{array}}} = \zeta''(\xi,
\xi).&
\end{eqnarray}
Let $\Gg_2$ be the group of power series $\ii + \sum_{\mathcal{T}}
f_\mathcal{T} Y^\mathcal{T}$. A straightforward adaptation of
proposition \ref{homomorphism} shows that
\begin{equation}
\label{hom2}
\Gg_2 \backsim G_2^{op}.
\end{equation}
Let us define two characters of $H_2$
\begin{equation}
\phi_{-}(\mathcal{T})\doteq \left\{
\begin{array}{l} \phi_1(\ttt) \text{ if } \mathcal{T} \in H_\bullet\\
                0 \text{ if }  \mathcal{T} \notin H_\bullet,
\end{array}\right.
\; , \quad  \phi_{+}(\mathcal{T}) \doteq \left\{
\begin{array}{l} \phi_1(\ttt) \text{ if } \mathcal{T} \in H_\ssquare\\
                0 \text{ if }  \mathcal{T} \notin H_\ssquare
\end{array}\right.
\label{phipm}
\end{equation}
where $H_\bullet$ is the set of trees decorated with bullets only
and $H_\ssquare$ those decorated by square only. Then
\begin{equation}
f_{\phi_1}[\zeta] = f_{\phi_{+}}[Y],\quad f_{\phi_1}[\xi] =
f_{\phi_-}[Y]
\end{equation}
are both elements of $G_2^{op}$. By (\ref{birkserie}) and
(\ref{hom2})
\begin{equation}
f_{\phi_1}[\chi]= f_{\phi}[Y]
\end{equation}
for the character $\phi\in G_2$ given by \footnote{ Remembering
(\ref{phiun}), $\phi_-^{-1}$ vanishes on all generators of $H_2$
except $\phi_-^{-1}(1) = 1, \;\phi_-^{-1}(\bullet) = -1$. By
(\ref{produit}) one obtains
\begin{equation}\label{phiexplcite}
\phi(\ttt) = \left\{\begin{array}{cl} 1 &\text{for } \ttt\in
H_\ssquare\\
-1 &\text{for } \ttt = \bullet\\
-n_\ttt &\text{otherwise}
\end{array}\right.
\end{equation}
where $n_\ttt = 1$ if there is a simple cut such that $
R_c(\ttt)\in H_\ssquare$ and $P_c(\ttt) = \bullet$, $n_\ttt = 0$
otherwise.}
\begin{equation}
\label{birkhoffdeux}
 \phi = \phi_{-}^{-1}* \phi_{+}.
\end{equation}
(\ref{birkserie}) can be written with a unique operator $Y$ and
three characters
\begin{equation}
\label{birkseriedeux} f_{\phi_+}[Y] = f_{\phi}  \circ
f_{\phi_-}[Y].
\end{equation}
(\ref{birkseriemoinsdeux}) can be written as well, by associating
bullet to $X_{1} - X_{0}$, square to $X_{1}$ and defining
$\phi_{\pm}$ according to $\phi_e$ instead of $\phi_1$.

(\ref{birkhoffdeux}) is the announced equation
(\ref{birkmoinsun}), establishing that the change of initial
condition in a fixed point equation is coded into the group of
characters of the algebra of decorated rooted trees. In next
section, we show that (\ref{birkhoffdeux}) (or equivalently
(\ref{birkseriedeux}) by identifying characters and series) allows
to define a so called algebraic Birkhoff decomposition.
%%%%%%%%%%%%%%%%%%%%%%%%%%%%%%%%%%%%%%%%%%%%%%%%%%%%%%%%%%%%%%%%%%%%%
%[{\bf I have to add that there is a work of Chapoton and Livernet
%that applies to a general context (pre-Lie algebras)}]
%%%%%%%%%%%%%%%%%%%%%%%%%%%%%%%%%%%%%%%%%%%%%%%%%%%%%%%%%%%%%%%%%%%%%%%
\subsection{Algebraic Birkhoff decomposition}

A decomposition of the same kind as (\ref{birkhoffdeux}) appears
in renormalization of quantum field theory with minimal
substraction scheme and dimensional regularization\cite{ck}: the
bare theory gives rise to a loop
\begin{equation}
\label{loop}
 \gamma(z) \in G_F, \quad z\in \mathcal{C}
\end{equation} where
$\mathcal{C}\in\cc$ is a small circle around the dimension $D$ of
space time and $G_F$ is the group of characters of the Hopf
algebra of Feynman diagrams $H_F$. The renormalized theory is the
evaluation at $z=D$ of the holomorphic part $\gamma_+$ of the {\it
Birkhoff decomposition} of $\gamma$. To give a similar
interpretation to our decomposition (\ref{birkhoffdeux}), we need
to adapt to rooted trees some of the tools of ref.[\citelow{ck}]
initially developed for Feynman diagrams.

An important feature is the commutative algebra $\aa$ of smooth
functions meromorphic inside $\mathcal{C}$ with pole only at $D$.
Also important are the subalgebras $\aa_+\subset\aa$ of functions
holomorphic inside $\mathcal{C}$, and $\aa_-\subset\aa$ the
subalgebra of polynomial in $\frac{1}{z-D}$ without constant term.
There exists a projection
\begin{equation}
p_-: \aa \rightarrow \aa_-
\end{equation}
parallel to $\aa_+$, i.e.
\begin{equation}
\label{ker} \text{ Ker } p_- = \aa_+.
\end{equation}
Feynman rules (ponderated by a suitable mass factor) yields an
algebra homomorphism
\begin{equation}
U: H_F \rightarrow \aa.
\end{equation}
The counterterms are given by the algebra homomorphism
\begin{equation}
C(X)  \doteq -p_-\lp U(X) + \sum C(X')U(X'')\rp
\end{equation}
where
\begin{equation}
\Delta(X) = X\ot 1 + 1\ot X + \sum X'\ot X'', \quad
X\in\tilde{H}_F\doteq\text{Ker } \epsilon,
\end{equation}
while the renormalized theory is given by the homomorphism
\begin{equation}
\label{ren} R(X) = (C*U)(X).
\end{equation}
One checks that
\begin{equation}
\label{curdeux} C(\tilde{H}_F)\subset\aa_-, \quad
R(\tilde{H}_F)\subset\aa_+.
\end{equation}
Equation (\ref{ren}) viewed as an equality between algebra
homomorphisms,
\begin{equation}
\label{cur}
 C* U = R,
\end{equation}
is called the {\it algebraic Birkhoff decomposition} of $U$. Such
a terminology is justified by considering the $G_F$-valued loops
\begin{equation}
\label{loopdeux} \gamma(z) \doteq \chi_z\circ U, \quad\gamma_-(z)
= \chi_z\circ C, \; \quad \gamma_+ (z) = \chi_z\circ R
\end{equation}
where
\begin{equation}
\chi_z(f) \doteq f(z) \quad \forall f\in\aa.
\end{equation}
Indeed (\ref{ren}) indicates that
\begin{equation}
\label{ckbirkhoff}
 \gamma_+(z) = \gamma_-(z)\gamma(z) \quad
\forall z\in\mathcal{C}
\end{equation}
for the pointwise product in $G$. One then shows that
(\ref{ckbirkhoff}) is precisely the Birkhoff decomposition of
$\gamma$. Namely, viewing $\mathcal{C}$ as a subset of the Riemann
sphere $\cc P_1$, $\gamma_-$ extends to a $G_F$-valued holomorphic
maps on $\mathcal{C_-}$ (the component of the complement of
$\mathcal{C}$ containing $\infty$) with $\gamma_-(\infty) = 0$
while $\gamma_+$ extends to $G_F$-valued holomorphic maps on
$\mathcal{C_+}$ (the other component of the complement of
$\mathcal{C}$). The reader is invited to consult
ref.[\citelow{ck}] for a precise definition of the Birkhoff
decomposition and its link to the Riemann-Hilbert problem. Let us
simply recall that the replacement of $\gamma$ by $\gamma_+$ is a
natural principle to extract a finite value from the singular
expression $\gamma(z)$.

The use of the algebra of meromorphic functions on $\mathcal{C}$
is intimately linked to dimensional regularization scheme. In the
framework of the continuous renormalization group, there is no
pole given by the dimension of space time. However, given a
decomposition of algebra homomorphisms such as (\ref{cur}),
satisfying conditions (\ref{curdeux}), it still makes sense to
talk of Birkhoff decomposition, but in the following algebraic
sense (taken from ref.[\citelow{kastler}] or [\citelow{ionescu}]).
{\defi \label{defibirkhoff} Let $H$ be a commutative Hopf algebra,
$\aa$ a commutative algebra with a projection $p_-$ on a
subalgebra $\aa_-$. An algebra homomorphism $\gamma: H\rightarrow
\aa$ has an algebraic Birkhoff decomposition if there exists two
algebras homomorphisms $\gamma_+$, $\gamma_-$ from $H$ to $\aa$
such that
\begin{equation}
\label{birkhofftrois}
 \gamma_+ = \gamma_-*\gamma
\end{equation}
where $*$ is the convolution product (\ref{produit}) and
\begin{equation}
\label{gammapm}
 p_+\gamma_+ = \gamma_+,\quad p_-\gamma_- =
\gamma_-
\end{equation} where $p_+$ is
the projection on
\begin{equation}
\label{aplus}
 \aa_+ = \text{Ker } p_-.
\end{equation}}

To interpret (\ref{birkhoffdeux})-(\ref{birkseriedeux}) as an
algebraic  Birkhoff decomposition one may be first tempted to
consider the algebra of polynomials in $Y^\ttt$ as the equivalent
in the continuous renormalization framework of the meromorphic
functions. As well characters may be understood as the equivalent
of the homomorphisms defined by Feynman rules: in the same way
that $U$, $C$, $R$ map a Feynman diagram to a meromorphic funtion,
characters map a decorated rooted tree to a monomial in $Y^\ttt$,
\begin{equation}
\label{algebrahom}
 \Phi(\mathcal{T}) \doteq \phi(\ttt)Y^\ttt,\;
\Phi_\pm(\ttt) \doteq \phi_\pm(\ttt)Y^\ttt.
\end{equation}
Unfortunately $\Phi$, $\Phi_\pm$ do not define an algebraic
Birkhoff decomposition. For instance (see the proof of proposition
\ref{birkprop} for the computation of $\phi$)
\begin{eqnarray}
\Phi_+ (\begin{array}{c}\ssquare\vspace{-3.55mm}\\ \vspace{-2.25mm} |\\ \circ\end{array}) &=& 0 \\
(\Phi_- * \Phi) (\begin{array}{c}\ssquare\vspace{-3.55mm}\\
\vspace{-2.25mm} |\\ \circ\end{array}) &=& \scl{\Phi_- \ot
\Phi}{1\ot \begin{array}{c}\ssquare\vspace{-3.55mm}\\
\vspace{-2.25mm} |\\ \circ\end{array} +
\begin{array}{c}\ssquare\vspace{-3.55mm}\\ \vspace{-2.25mm} |\\
\circ\end{array}\ot 1 +
\bullet\ot\ssquare}\\
\label{essaibirk} &=& -Y^{\begin{array}{c}\ssquare\vspace{-3.55mm}\\
\vspace{-2.25mm} |\\ \circ\end{array}} + Y^{\circ} Y^{\ssquare}.
\end{eqnarray}
A solution would be to define the product of monomials so that
(\ref{essaibirk}) vanishes,
\begin{equation}
Y^{\ssquare}Y^{\circ} =
Y^{\begin{array}{c}\ssquare\vspace{-3.55mm}\\ \vspace{-2.25mm} |\\
\circ\end{array}}
\end{equation}
but such a product would not be commutative. However by
considering the (commutative) formal product of decorations, it
appears that the sum (ponderated by the numerical coefficient) of
the products of decorations in (\ref{essaibirk}) vanishes (i.e.
$-\circ\ssquare + \circ\ssquare = 0$). Consequently we propose a
Birkhoff decomposition with value on the the algebra $\aa$ of
decorations, that is to say the free unital algebra generated by
square and bullet
\begin{equation}
\label{alg} \aa = \overline{\left\{1, \bullet, \ssquare \right\}}.
\end{equation}
For $\aa_-$ we choose the unital subalgebra of $\aa$ generated by
the bullet
\begin{equation}
\aa_- = \overline{\left\{ 1, \bullet\right\}}.
\end{equation}
We note $p_-$ the projection $\aa \rightarrow \aa_-$
\begin{equation}
p_-(1)= 1,\quad p_-(\bullet)= \bullet,\quad p_-(\ssquare)= 0
\end{equation}
extended to all $\aa$ by algebra homomorphism
\begin{equation}
p_-(\bullet^n\, \ssquare^m + \bullet^{n'}\, \ssquare^{m'}) =
\bullet^n + \bullet^{n'}.
\end{equation}
$\aa_+$ and $ p_+$ are defined by (\ref{aplus}). Let $\Gamma$ be
the algebra homorphism $H_2 \rightarrow \aa$
\begin{equation}
\Gamma(1)= 1,\quad \Gamma(\mathcal{T}) =
\bullet^{\abs{\mathcal{T}}_\circ}\,
\ssquare^{\abs{\mathcal{T}}_\square}
\end{equation}
where $\abs{\mathcal{T}}_\circ$ is the number of bullets of $\ttt$
and $\abs{\mathcal{T}}_\square$ is the number of squares. $\Gamma$
just "counts the decorations", for instance
\begin{equation}
\Gamma ({\begin{array}{c}
\ssquare\vspace{-3.3mm}\\
\wedge\vspace{-3.3mm}\\
\circ\,\circ\vspace{-3.6mm}\\ \hspace{2.6mm}\shortmid\vspace{-3.3mm}\\
\hspace{2.6mm}\circ
\end{array}}) = \bullet^3\ssquare.
\end{equation}
Finally we define three algebra homomorphisms from $H_2$ to $\aa$,
\begin{equation}
\label{algebrahom}
 \gamma(\mathcal{T}) = \phi(T)\Gamma(T),\;
\gamma_\pm(\mathcal{T}) = \phi_\pm(T)\Gamma(T).
\end{equation}

{\prop \label{birkprop} $\gamma_+ = \gamma_- * \gamma$ is an
algebraic Birkhoff decomposition.} \newline

\noindent {\it Proof.}  Let $\overline{H_\bullet}$ be the algebra
of trees with bullets only (and similarly
$\overline{H_\ssquare}$). Then
\begin{equation}
\gamma_-(\ttt) =\left\{
\begin{array}{ll}
\bullet^{\abs{\ttt}_\circ} &\text{for } \ttt\in \overline{H_\bullet}\\
0 &\text{otherwise,}
\end{array}\right.
\quad \gamma_+(\ttt) =\left\{
\begin{array}{ll}
\ssquare^{\abs{\ttt}_\square} &\text{for } \ttt\in \overline{H_\ssquare}\\
0 &\text{otherwise,}
\end{array}\right.
\end{equation}
so that (\ref{gammapm}) is satisfied, as well as (\ref{aplus}) by
construction. By algebra homomorphism, (\ref{birkhofftrois}) has
to be checked only on generators. Let us first note that
\begin{equation}
\label{gammatc}
 \gamma(\ttt) =\left\{
\begin{array}{ll}
\ssquare^{\abs{\ttt}_\square} &\text{for } \ttt\in H_\ssquare\\
- \bullet &\text{for } \ttt = \bullet\\
-n_\ttt\, \bullet^{\abs{\ttt}_\circ}\,
\ssquare^{\abs{\ttt}_\square} &\text{otherwise}
\end{array}\right.
\end{equation}
where $n_\ttt$ is defined in (\ref{phiexplcite}).
 Second, for any $\ttt\neq 1$
\begin{equation}
(\gamma_-*\gamma)(\mathcal{T}) = \gamma(\mathcal{T}) +
\gamma_-(\mathcal{T})+ \sum_{c\in C(\mathcal{T})}
\gamma_-(P_c(\mathcal{T})) \gamma(R_c(\mathcal{T})).
\end{equation}
This is then not difficult to check (\ref{birkhofftrois}) for $
\ttt\in H_\bullet$ or $\ttt\in H_\ssquare.$

For $\ttt\notin H_\bullet \cup H_\ssquare$,
\begin{equation}
\label{birkquatre} (\gamma_-*\gamma)(\mathcal{T}) = -n_\ttt\,
\bullet^{\abs{\ttt}_\circ} \, \ssquare^{\abs{\ttt}_\square} +
\sum_{c\in C(\mathcal{T})} \gamma_-(P_c(\mathcal{T}))
\gamma(R_c(\mathcal{T})).
\end{equation}
Assume $n_\ttt =0$. Then (\ref{birkquatre}) is non zero only if
there is at least either a simple cut $c$ such that
\begin{equation}
P_{c}(\ttt)\in\overline{H_\bullet} \text{ and } R_{c}(\ttt) \in
H_\ssquare
\end{equation}
 or a simple
cut $\tilde{c}$ with
\begin{equation}
P_{\tilde{c}}(\ttt)\in\overline{H_\bullet} \text{ and }
n_{R_{\tilde{c}}(\ttt)}\neq 0.
\end{equation}
Assume there exists a simple cut $c$ for which $P_c(\ttt)$ is a
single tree. Then there exist a $\tilde{c}$ in which
$R_{\tilde{c}}(\ttt)$ is the subtree of $\ttt$ consisting in
$R_c(\ttt)$ and the root of $P_c(\ttt)$ while
$P_{\tilde{c}}(\ttt)$ is the union of all the subtrees of
$P_c(\ttt)$ obtained by promoting as roots the vertices of length
$1$ ($P_{\tilde{c}}(\ttt)\neq \emptyset$ because $n_\ttt = 0$).
The simple cut $c$ contributes to (\ref{birkquatre}) with a factor
\begin{equation}
\label{contrun}
 \bullet^{\abs{P_c(\ttt)}}  \ssquare^{\abs{R_c(\ttt)}}
\end{equation}
whereas $\tilde{c}$ contributes with a factor
\begin{equation}
\label{contrdeux} \bullet^{\abs{P_{\tilde{c}}(\ttt)}} (-
\bullet\ssquare^{\abs{R_{\tilde{c}}(\ttt)}_\square}).
\end{equation}
This is easy to observe that (\ref{contrun}) = -
(\ref{contrdeux}). The same is true if $P_c(\ttt)$ is a product of
$m$ trees, the only difference being that the contribution of $c$
is cancelled by the sum of the contributions of the $m$
$\tilde{c}$ obtained by grafting alternatively to $R_c(\ttt)$ each
of the roots of $P_c(\ttt)$. Also, starting from a $\tilde{c}$,
one would similarly notice that its contribution is cancelled by a
$c$ so that, finally, (\ref{birkquatre}) vanishes for any
$\ttt\notin H_\ssquare\cup H_\bullet$ with $n_\ttt=0$. Hence
(\ref{birkhofftrois}).

The same procedure applies when $n_\ttt\neq 0$. The term of
(\ref{birkquatre}) in $\gamma(\ttt)$ is cancelled by the sum of
the $n_\ttt$ terms corresponding to the simple cuts with
$P_c(\ttt) = \bullet$. \hfill $\blacksquare$
\newline

As announced in (\ref{birkmoinsun}), we have shown that the change
of initial condition in a fixed point equation does correspond to
the algebraic Birkhoff decomposition of the algebra morphisms
associated by (\ref{algebrahom}) to the characters encoding the
solutions. By convention we say that $\gamma_+$ is the positive
part of the algebraic $\aa$-valued Birkhoff decomposition of
$\gamma$.

\subsection{Continuous renormalization group}

Birkhoff decomposition has been introduced in renormalization of
quantum field theory by Connes and Kreimer in the framework of
minimal substraction scheme and dimensional regulari\-zation. Our
algebraic Birkhoff decomposition (\ref{birkhoffdeux}) has an
interpretation in the framework of the continuous renormalization
group of Wilson\cite{wilson} (see also ref.[\citelow{morris}] for
a nice introduction). The latest describes the evolution of the
parameters of a quantum field theory under a change of the
observation scale. When the rescaling is parametrized by a
continuous quantity, say the energy, the evolution is governed by
flow equation
\begin{equation}
\label{dsdlambda} \Lambda \frac{\partial}{\partial \Lambda} S =
\beta(\Lambda, S)
\end{equation}
 in which $\Lambda\in\rr^{*+}$ is the scale
and $S=S(\Lambda)$ implements the parameters (mainly $S$ is a
functional of the fields and their coupling constants). The
initial conditions are encoded by the knowledge of $S_0\doteq
S(\Lambda_0)$ at a given scale $\Lambda_0$. A typical example of
continuous renormalization group equations (\ref{dsdlambda}) are
Polchinski's equations \cite{polchinski} which gives
$\beta(\Lambda, S)$ for a theory with IR cutoff.

Let us consider the most general context by simply assuming that
the theory is described by a smooth operator $S$
$$S: \Lambda \mapsto S(\Lambda)\in\ee$$
where $\ee\doteq\ss(\hh)$ and $\hh$ is the (infinite dimensional)
vector space spanned by vectors labelled with the parameters of
the theory. A scale transformation
$$\Lambda \rightarrow \Lambda^s$$
with $s\in\rr^{*+}$ induces
 the transformation \cite{bellac}
$$S(\Lambda) \rightarrow S(\Lambda^s) = s^\Delta S(\Lambda)$$
where $\Delta\in\ee$ is the diagonal matrix whose coefficients are
the dimensions of the parameters. We define the dimensionless
quantities $x(\Lambda)\in\ee$ and $t\in\rr$,
$$S(\Lambda) = \Lambda^\Delta x(\Lambda)\, , \quad t\doteq \ln(\frac{\Lambda}{\mu}) $$
where $\mu$ is a parameter of the theory with same dimension as
$\Lambda$.  Then (\ref{dsdlambda}) yields
\begin{equation}
\label{evoldeux}
 \La \frac{\partial x}{\partial \La}= -\Delta x +
\La^{-\Delta}  \beta(\La,\La^{\Delta} x ).
\end{equation}
By dimensional analysis, $\beta$ is transforming homogeneously
under change of scale,
\begin{equation}
\beta(\La,\La^{\Delta} x) =\La^{\Delta} \beta(1,x)
\end{equation}
so that, writing $\hbar X(x) \doteq \beta(1,x)$ and $D= -\Delta$,
\begin{equation}\label{equacste}
\frac{\partial x}{\partial t} = D x + \hbar X(x)
\end{equation}
whose integral form, with initial condition $x_0 = \Lambda_0^{D}
S_0$, is
\begin{equation}
\label{polchtrois}
 x(t) = e^{(t-t_0)D} x_0 + \hbar\int_{t_0}
^t e^{(t-u)D} X(x(u))du.
\end{equation}
Viewing $x$ as an element of $\ee'$, Banach vector space of
applications from $\rr^{*+}$ to $\ee$,  we define
\begin{equation}
\label{xtilde}
 \tilde{x}_0\in\ee': t \mapsto  e^{(t-t_0)D} x_0
\end{equation}
and $\chi\in \ss(\ee')$,
\begin{equation}\chi(x): t \mapsto \int_{t_0}
^t e^{(t-u)D} X(x(u))du \quad \forall x\in\ee' \end{equation} so
that (\ref{polchtrois}) reads as the fixed point equation
\begin{equation}
\label{contun}
 x = \tilde{x_0} + \hbar\chi(x).
\end{equation}

In the context of Wilson's continuous renormalization group,
$\Lambda_0$ is interpreted as an UV cutoff and one is interested
in the limits of very high energy scale, i.e. $t_0
\rightarrow+\infty$. However $\tilde{x}_0$ is already ill-defined
since
\begin{equation} e^{(t-t_0)D} x_0 \left\{
\begin{array}{l}
\text{ converges on } \hh^{+} \\
\text{ is constantly zero on } \hh^{0} \\
\text{ diverges on } \hh^{-}
\end{array} \right. \text{ as $t_0\rightarrow +\infty$}
\end{equation}
where $\hh^{+}$, $\hh^{0}$, $\hh^{-}\subset \hh$ are the proper
subspaces of $D$ corresponding to positive, zero and negative
eigenvalues (the corresponding parameters are called {\it
relevant}, {\it marginal} and {\it irrelevant}). $\chi$ as well is
ill defined, even in the relevant sector. A solution to ensure the
finiteness of $x(t)$ at high scale consists in fixing the initial
conditions for the irrelevant sector at scale $t_1$ (equivalently:
at scale $\Lambda_1$) distinct from $t_0$. Namely one imposes the
mixed boundary conditions
\begin{equation}x_{R} \doteq P \tilde{x}_1 + (\ii-P) \tilde{x}_0 \end{equation}
where $P$ is the orthogonal projection from $\hh$ to $\hh^-$ and
\begin{equation} \label{xtildeun} \tilde{x}_1\in\ee': t \mapsto  e^{(t-t_1)D}
x_1
\end{equation}
Defining
\begin{equation}
\rho(x): t \mapsto \int_{t_1} ^t e^{(t-u)D} X(x(u))du
\end{equation} and
\begin{equation}
\zeta\doteq P\rho +(\ii - P)\chi
\end{equation}
allows to write (\ref{contun}) with initial condition $x_R$
\begin{equation}
\label{contdeux}
 x(t)  = x_{R} + \hbar\zeta(x).
\end{equation}

This is a well known result that $x(t)$ computed with mixed
boundary condition, namely
\begin{equation}
x(t) = f_{\phi_1}[\zeta](x_R)
\end{equation}
remains finite at high energy scale and does not depend on $x_0$.
Trees have already been used to prove this result (see
ref.[\citelow{hurd}] for instance). To be complete we propose here
a simple proof taking advantage of the Hopf algebraic structure.

 {\prop
$\underset{t_0 \rightarrow +\infty}{\lim} f_{\phi_1}[\zeta](x_R)$
is finite order by order and does not depend on $x_0$.}
\newline

\noindent {\it Proof.} The idea is to use decorated trees in a
similar way as (\ref{y}) except that the rule now is
\begin{equation}
Y^\bullet = \zeta_1 \doteq P\rho,\quad Y^\ssquare = \zeta_2 \doteq
(\ii -P)\chi.
\end{equation}
Then
\begin{eqnarray}
\zeta^{\small{\begin{array}{c}\vspace{-3mm}\bullet\\ \vspace{-2.75mm}\shortmid\\
\circ\end{array}}}
&=& (\zeta_1 + \zeta_2)'(\zeta_1 + \zeta_2) \\
&=& (\zeta_1)'(\zeta_1) + (\zeta_1)'(\zeta_2) +
(\zeta_2)'(\zeta_1) + (\zeta_2)'(\zeta_2)\\
&=&
Y^{\small{\begin{array}{c}\vspace{-3mm}\bullet\\ \vspace{-2.75mm}\shortmid\\
\circ\end{array}}}+
Y^{\small{\begin{array}{c}\vspace{-3mm}\bullet\\ \vspace{-1.95mm}\shortmid\\
\square\end{array}}} +
Y^{\small{\begin{array}{c}\vspace{-3.2mm}\ssquare\\ \vspace{-2.6mm}\shortmid\\
\circ\end{array}}} +
Y^{\small{\begin{array}{c}\vspace{-3mm}\ssquare\\ \vspace{-1.95mm}\shortmid\\
\square\end{array}}}\\
&=& \sum_{\ttt\in\left\{{\small{\begin{array}{c}\vspace{-3mm}\bullet\\ \vspace{-2.75mm}\shortmid\\
\circ\end{array}}}\right\}_{2}} Y^\ttt
\end{eqnarray}
where $\left\{T\right\}_{2}$ is the set of all decorated trees
obtained by decoration of the vertices of $T\in H$, e.g.
\begin{equation}
\left\{\bullet \right\}_{2} = \left\{\bullet, \ssquare \right\}.
\end{equation}
It is not difficult to check that the same is true for any $T\in
H$,
\begin{equation}
\zeta^{T}= \sum_{\ttt\in\left\{ T \right\}_{2}} Y^\ttt.
\end{equation}
Thus
\begin{equation}
f_{\phi_1}[\zeta]= \sum_{\ttt \in H_2} Y^\ttt.
\end{equation}
Since
\begin{equation}
\label{xrrr}
\underset{t_0 \rightarrow +\infty}{\lim} Y^1(x_R)  = \underset{t_0
\rightarrow +\infty}{\lim} x_R = \tilde{x}_1
\end{equation}
and $Y^\bullet$
does not depend on $t_0$ then
\begin{equation}
\label{convergun}
 \underset{t_0 \rightarrow +\infty}{\lim}
Y^\bullet(x_R) = Y^\bullet(\tilde{x}_1).
\end{equation}
Similarly for any $\ttt\in H_\bullet$, $\underset{t_0 \rightarrow
+\infty}{\lim} Y^\ttt(x_R)$ is both $x_0$ and $t_0$ independent.
In the same way
\begin{equation}
 \label{convergdeux}\underset{t_0
\rightarrow +\infty}{\lim} Y^\ssquare(x_R) =  \underset{t_0
\rightarrow +\infty}{\lim} (\ii-P)\int_{t_0} ^t e^{(t-u)D}
X(\tilde{x}_1(u))du
\end{equation}
is finite because $X$ is smooth. Similarly for any $\ttt\in
H_{\text{\tiny $\ssquare$}}$. For $\ttt$ decorated with both
bullets and squares, combination of (\ref{convergun}) and
(\ref{convergdeux}) ensures that $Y^\ttt$ is finite and $x_0$
independent.\hfill $\blacksquare$
\newline

Now, using decorated rooted trees with the rule (\ref{y}), one
identifies $f_{\phi_1}[\zeta](x_R) = f_{\phi_+}[Y](x_R)$ as the
renormalized theory, $f_{\phi_1}[\chi](x_R) = f_{\phi}[Y](x_R)$ as
the bare theory and the counterterms are given by
$f_{\phi_-}[Y](x_R)$. By proposition \ref{birkprop} one finally
obtains the main result of this paper:
\newline

{\it The bare and renormalized theories define two algebra
morphisms $\gamma$ and $\gamma_+$ between the Hopf algebra of
decorated rooted trees and the free algebra $\aa$ of decorations.
$\gamma_+$ is the positive part of the algebraic $\aa$-valued
Birkhoff decomposition of $\gamma$.}

\section{Conclusion}

We have presented the first algebraic steps towards an adaption of
Connes-Kreimer work to the ERGE. As in the BPHZ procedure with
minimal substraction scheme in dimensional regularization,
renormalized and bare theories are linked inside a Birkhoff
decomposition. In the BPHZ framework, renormalization corresponds
to the projection of meromorphic functions (with pole only at
dimension of space time) on their holomorphic part. Continuous
renormalization appears as a projection on one decoration. Whether
the algebra of decorations is an artefact -hiding deeper
connection with the Rieman-Hilbert problem - or is truly
meaningful is not perfectly clear at the moment. This will be the
object of further investigations, as well as the precise relation
between trees and Feynman diagrams and their use as computational
tools for effective actions.
\newline

\subsection*{Acknowledgment} P.M. is supported by the european
network "Geometric analysis".


\begin{thebibliography}{98}
\bibliographystyle{plain}
\markboth{\uppercase{Bibliographie}}{\uppercase{Bibliographie}}
\addcontentsline{toc}{section}{Bibliographie}
\labelsep 0pt
\bibitem{brouder} C. Brouder, {\it Runge-Kutta methods and
renormalization}, Eur. Phys. J. C {\bf 12} (2000) 521-534.
\bibitem{butcher} J. C. Butcher, {\it An algebraic theory of
integration methods}, Math. Comput. {\bf 26} (1972) 79-106.
\bibitem{cayley} A. Cayley, {\it On the theory of analytical forms
called trees} Phil. Mag {\bf 13} (1857) 172-6.
\bibitem{ck} A. Connes,  D. Kreimer,  {\it Hopf algebras, renormalization and noncommutative geometry},
Comm.Math.Phys. {\bf 199} (1998) 203 A. Connes, D. Kreimer,  {\it
Renormalization in quantum field theory and the Riemann-Hilbert
problem I: the Hopf algebra structure of graphs and the main
theorem}, Commun. Math. Phys. {\bf 210} (2000) 249 A. Connes, D.
Kreimer, {\it Renormalization in quantum field theory and the
Riemann-Hilbert problem II: the $\beta$-function, diffeomorphisms
and the renormalization group}, Comm. Math. Phys. {\bf 216} (2001)
215
\bibitem{fig} H. Figueroa, J.M. Gracia Bondia, {\it On the antipode of Kreimer's Hopf algebra}, hep-th/9912170, San José, 1999.
\bibitem{gallavotti} G. Gallavotti, F. Nicolo, {\it Renormalization theory in four dimensional scalar fields 1},
Comm. Math. Phys. {\bf 100} (1985) 545,  {\it Renormalization
theory in four dimensional scalar fields 2}, Comm. Math. Phys.
{\bf 101} (1985) 247
\bibitem{jgb}  J. M. Gracia-Bondia, J. C. Varilly, H. Figueroa, {\it
Elements of noncommutative geometry}, Birkhauser (2001).
\bibitem{wanner} E. Hairer, G. Wanner {\it On the Butcher grouup and general multi-value methods}, Computing {\bf
13}, 1-15 (1974)
\bibitem{hurd}  T. Hurd, {\it A renormalization group proof of perturbative renormalizability}, Comm. Math. Phys. (1989)
\bibitem{ionescu} L. M. Ionescu, M. Marsalli, {\it A Hopf algebra
deformation approach to renormalization}, hep-th/0307112.
\bibitem{itzub} C. Itzykson, J.B. Zuber, {\it Quantum field theory}, McGraw-Hill
(1985).
\bibitem{bellac} M. Le Bellac, {\it Des phénomènes critiques aux
champs de jauge}, InterEditions (1990).
\bibitem{kastler} D. Kastler, {\it Connes-Moscovici-Kreimer Hopf algebras},  Fields Institute Communications {\bf 30} 2001.
\bibitem{kogut} J. Kogut, K. Wilson,  {\it The renormalization group and the $\epsilon$-expansion},  Phys. Rept. 12 (1974) 75
\bibitem{gkm} F. Girelli, T. Krajewski, P. Martinetti, {\it Wave-Function renormalization and the
Hopf algebra of Connes and Kreimer}, Mod.Phys.Lett. {\bf A16}
(2001) 299-303.
\bibitem{morris}  T. Morris, {\it Elements of the continuous renormalization group}, hep-th/9802039
\bibitem{peterman}  A. Petermann, E. Stueckelberg,  {\it  The renormalization group in quantum theory} Helv.Phys.Acta 24 (1951) 317
\bibitem{polchinski}  J. Polchinski, {\it Renormalization and effective lagrangians}, Nucl. Phys. B 231 (1984) 269
\bibitem{rivasseau} V. Rivasseau, {\it From perturbative to constructive renormalisation}, Princeton University Press (1991)
\bibitem{wilson} K. G. Wilson, {\it Renormalization group
methods}, Adv. Math. {\bf 16} (1975) 170-186.
\end{thebibliography}
\end{document}